# Diversity in the radiation-induced transcriptomic temporal response of mouse brain tissue regions

Karolina Kulis[1], Sarah Baatout[2], Kevin Tabury[2], Joanna Polanska[1*], Mohammed Abderrafi Benotmane[2]

[1] Department of Data Science and Engineering, Silesian University of Technology, Gliwice, Poland
[2] Radiobiology Unit, Institute for Nuclear Medical Application, Belgian Nuclear Research Centre, Belgium

* Corresponding author: joanna.polanska@polsl.pl

ABSTRACT

A number of studies have indicated a potential association between prenatal exposure to radiation and late mental disabilities. This is believed to be due to long-term developmental changes and functional impairment of the central nervous system following radiation exposure during gestation. This study conducted a bioinformatics analysis on transcriptomic profiles from mouse brain tissue prenatally exposed to increasing doses of X-radiation. Gene expression levels were assessed in different brain regions (cortex, hippocampus, cerebellum) and collected at different time points (at 1 and 6 months after birth) for C57BL mice exposed at embryonic day E11 to varying doses of radiation (0, 0.1 and 1 Gy). This study aimed to elucidate the differences in response to radiation between different brain regions at different intervals after birth (1 and 6 months). The data was visualised using a two-dimensional Uniform Manifold Approximation and Projection (UMAP) projection, and the influence of the factors was investigated using analysis of variance (ANOVA). It was observed that gene expression was influenced by each factor (tissue, time, and dose), although to varying degrees. The gene expression trend within doses was compared for each tissue, as well as the significant pathways between tissues at different time intervals. Furthermore, in addition to radiation-responsive pathways, Cytoscape's functional and network analyses revealed changes in various pathways related to cognition, which is consistent with previously published data [1] [2] [3], indicating late behavioural changes in animals prenatally exposed to radiation

KEYWORDS

Radiation, brain, mouse, microarray, bioinformatics,

# INTRODUCTION

Epidemiological data from in utero exposed Hiroshima and Nagasaki A-bomb survivors indicates that prenatal irradiation may severely compromise normal embryonic development, leading to a variety of anatomical, physiological and mental defects. The assessment of radiation exposure can vary for different organisms and cell types involved in radiobiology experiments. Advances have been made in animal models, such as rodents, to explore the effects of radiation on the unborn child. These models have been used to reproduce these effects in order to better understand the underlying cellular and molecular mechanisms. Most research agrees on the idea that the brain is sensitive to radiation at the prenatal or childhood stage. This is presumably due to incomplete developmental processes. A number of authors have reported on the effects of radiation on learning and memory in various types of mazes, including [1], [2], [3], [7], [8], [9], [10]. Despite its therapeutic benefit, radiation therapy is associated with a number of adverse effects, including cognitive and emotional dysfunctions and changes in neuroanatomical structures, particularly in young patients [11], [12]. It is therefore of great importance to assess the impact of radiation on living organisms and to identify markers that allow for the assessment of the absorbed dose. This article presents a bioinformatics analysis of the potential effects of radiation exposure during early gestation on gene expression in adulthood. The study aims to investigate differences in response to irradiation between brain regions by examining whether there are robust, tissue-, time-, and dose-independent biomarkers of irradiation and comparing gene expression values for different doses. The objective of this study is to investigate the differences in response to irradiation between brain regions. In addition, the study aims to identify robust, tissue-, time-, and dose-independent biomarkers of irradiation and to compare gene expression levels for different doses.

# MATERIAL AND METHODS

*Animals and irradiation*

All animal experiments were performed in accordance with the European Communities Council Directive (2010/63/EU) and approved by the local ethical SCK•CEN (Brainres) or SCK•CEN/VITO (ref. 02–012 and 11-005), University of Antwerp and KU Leuven committees. C57Bl/6J were purchased from Janvier (Bio Services, Uden, The Netherlands) or from Charles River Breeding Laboratories (Leiden, The Netherlands). Animals were housed under standard laboratory conditions (12-h light/dark cycle) and food and water were available *ad libitum*. Female mice were coupled during a 2-h time period in the morning, at the start of the light phase, in order to ensure synchronous timing of embryonic development. Subsequently, dams were whole-body irradiated at E11 (0.10 or 1.00 Gy) at a dose rate of 0.35 Gy min−1 using a Pantak RX tube operating at 250 kV and 15 mA (1-mm Cu-filtered X-rays). The calibration of this X-ray tube was performed using an ionisation chamber. For all irradiation procedures, mice were placed in a plexiglass pie-shaped box, in which their movement was restricted. Subsequently, the plexiglass box containing the animals was transported to the irradiation facility and placed within the radiation field, ensuring an equal dose distribution to all mice. After the irradiation, mice were placed back (returned) to their home cage. Control mice were sham-irradiated, i.e. they underwent all procedures similar to irradiated animals but were not placed within the radiation field.

*Microarray-based transcriptomic profile measurement*

Total RNA was extracted from flash-frozen adult brain tissues (Cortex, hippocampus and cerebellum) at 1 or 6 months after birth using the AllPrep DNA/RNA/Protein Mini Kit (Qiagen, Hilden, Germany) and quality-controlled using the 2100 BioAnalyzer (Agilent, Santa Clara, CA, US). Only samples with an RNA integrity number >8 were used for hybridisation onto Affymetrix Mouse Gene 2.0 ST arrays (Affymetrix, Santa Clara, CA, US) as per the manufacturer's recommendations. CEL files were uploaded for subsequent bioinformatics analyses. Data normalisation was performed using a customised Robust Multi-array Average algorithm (background correction for entire probe sequence, quantile normalisation, log2 transformation of intensity signals).

*Statistical analysis*

Statistical analysis was performed using NumPy, Pandas, Scikit-learn, Statsmodels, and Matplotlib, in Python version 3.10. Descriptive statistics, including minimum and maximum values, mean, median, standard deviation, lower and upper quartiles, and interquartile range, were calculated for each variable and dose/timepoint, along with their corresponding 95% confidence intervals. The internal data structure was visualised and investigated using the UMAP [13] transformation. To achieve this, the PCA high-dimensionality reduction technique was used to select the first components that explain 95% of the total variance in the dataset. Subsequently, the data was transformed using the 2D UMAP transformation with the new PCA-based set of features and Euclidean similarity measure. Outliers were checked using Dixon's Q test. A three-way analysis of variance (ANOVA) test was performed to identify differentially expressed genes, with tissue, time, and dose as independent variables. The general effect size was estimated as eta squared [14] and modified Glass delta estimates supported post hoc pairwise comparisons. The analyses were summarised using Venn diagrams. The further study investigated the dose trend for genes that did not exhibit significant interactions between factors across various brain regions and time intervals following exposure to irradiation.

*Functional analysis*

Functional analysis using Gene Set Enrichment Analysis (GSEA) [15] [16] was performed to identify radiation-enriched signalling pathways. It was done mostly in R programming language version 4.3.1 and Bioconductor's *fgsea* package. All genes from the original dataset were used and ranked according to the effect size value of the dose factor [14]. MSigDB database was used to gather only the gene sets related to irradiation and brain development. Additionally, we applied a filter to the database of gene sets for mice species using a set of keywords to be found in the pathway descriptions. The keywords we used were *radiation*, *irradiation*, *cognitive function*, *brain development*, *neurological response*, *neurogenesis*, *memory*, *cognition*, *behaviour*, and *learning*. The signalling pathway filtration procedure resulted in a selection of 860 gene sets that were considered in our functional analysis. The gene sets resulting from the GSEA were carefully curated to ensure that only pathways relevant to the project's aim were included. Pathways that were not associated with neurological functions were excluded. For further details, please refer to the supplementary materials (Table S3-S8, Figure S1-S6). The relationship networks between gene sets with an enrichment p-value < 0.05, which were considered significantly enriched, were jointly analyzed using Cytoscape software after calculating the similarity coefficient.

# RESULTS

Gene expression levels of 27,352 transcripts, assessed in different brain regions (cortex, hippocampus, cerebellum) and collected at different time points (at 1 and 6 months after birth) of C57bl mice exposed during gestation at embryonic day E11 to different doses of radiation (0, 0.1 Gy and 1 Gy) constituted the data.

PCA technique, with choosing the components explaining 95% of the total variance in the dataset, reduced the dimension from 27,352 to 53. The 2D UMAP transformation was then applied, resulting in a plot shown in Figure 1. To make the visualisation clearer and more legible, the differentiating factors were distinguished as follows: different doses were differentiated on the plot with colour, time with size and tissue with shape. One sample was tested positive for being an outlier using Dixon's Q test and removed from further analysis.

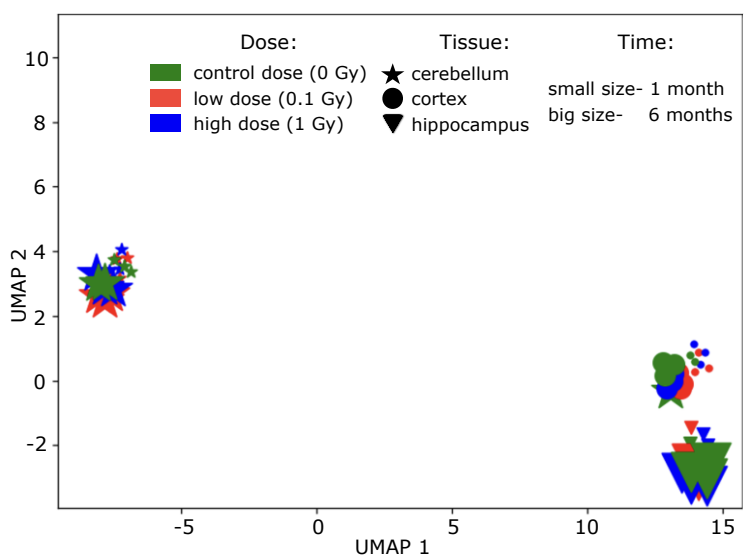

*Figure 1. Data visualised- UMAP transformation with differentiation by factors.*

Three-way Analysis of Variance (ANOVA), where the independent variables were tissue, time and dose, while the response variable was the gene expression, was performed and supported by the eta-squared effect size measures. The genes with at least a large effect size for all factors (main effects separately and interactions between them) were summarised in the Venn diagram (see Figure 2).

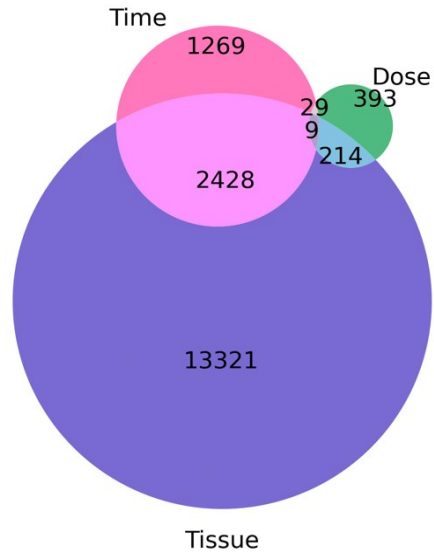

*Figure 2. Venn diagram of large effects (numbers within each circle correspond to the number of genes fulfilling the selection criterion)*

Performing ANOVA has proved that each factor (tissue, time and dose) has influenced gene expression, although each to a different degree. The tissue factor differentiated the most and time the least, which was already noticeable after applying UMAP transformation to the data. 13,321 transcripts were tissue-, 1,269 time-, and 393 were dose-dependent only with at least a large effect size. 29 transcripts were time- and dose-dependent, 214 dose- and tissue-dependent and 2428 time- and tissue-dependent. 9 genes had a large effect value for all factors.

The gene expression time series were investigated to check how gene expression reacted after the irradiation. From the 393 dose-dependent genes, the ones that had no significant interactions between the differentiating factors were considered (see supplementary materials). To demonstrate the impact of the interaction on the gene expression response, Figure 3A presents an example of a gene with inter-factor interactions. Figure 3B shows a case where the interactions are not significant – tissue-specific time responses seem to be similar.

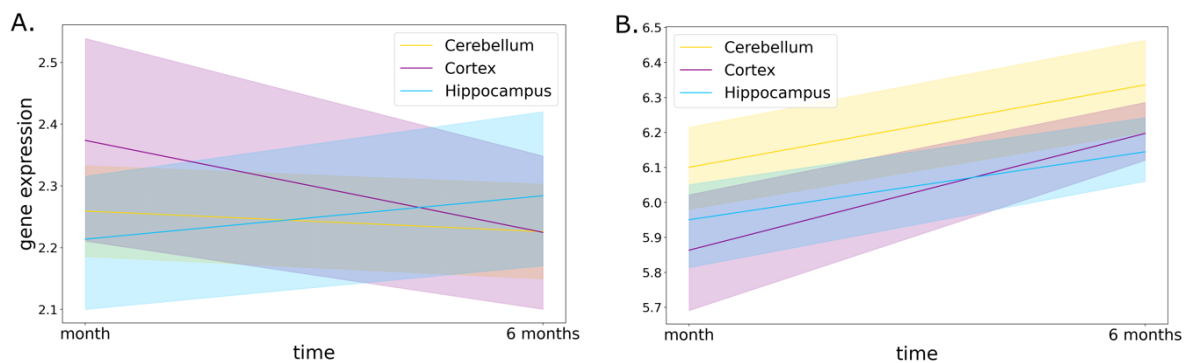

*Figure 3. A. Interaction plot of a gene with significant interactions between factors (left panel) tissue dependent different time responses B. Interaction plot of a gene with no significant interactions between factors (right panel). The solid lines represent the approximated time response for different tissue samples, while the colour-coded shadowed area visualises their 95% confidence intervals.*

A similar trend analysis was performed for the dose series (for the 393 genes) within various brain regions and in different time intervals after the exposure of irradiation. The constant increase of gene expression (control dose < low dose < high dose) or constant decrease in dose (control dose > low dose > high dose), independently of time point, is an example of dose dependency. The examples of boxplots presenting such a behaviour can be seen in Figure 4 (panels A and B), where the yellow line presents the median of gene expression for different doses. At one month after exposure to irradiation, a constant increase in dose was observed in 52 genes for the cerebellum, 54 genes for the cortex and 50 genes for the hippocampus, while the constant decrease was present in 41 genes for the cerebellum, 46 genes for the cortex and 49 genes for the hippocampus (see Table 1). After six months of irradiation, the increasing trend was observed in 52 genes for the cerebellum tissue, 51 genes for the cortex and 72 genes for the hippocampus, while the decreasing trend was present in 57 genes for the cerebellum, 53 genes for the cortex and 44 genes for the hippocampus (see Table 2). The full list of genes [17] obtained in each group can be studied in the supplementary materials (Tables S1 and S2). A literature review of the genes [18] obtained in each group was performed to identify the genes that are functionally most related to the effect of radiation on neurological and cognitive functions in different brain tissues. The ascending trend in the tissues was observed among others in Tti2 (encodes a regulator of the DNA damage response; the protein is a component of the Triple T complex (TTT), which also includes telomere length regulation protein and TELO2 interacting protein 1) for cerebellum after one month of irradiation and for hippocampus after both one and six months after the exposure. In the cerebellum, this trend was also present in the Tcf4 gene (encodes transcription factor 4, a basic helix-loop-helix transcription factor) independently of the time factor. After six months of irradiation, the same pattern in dose was observed in Ecrg4 (Esophageal Cancer Related Gene 4 Protein enables neuropeptide hormone activity; involved in neuropeptide signalling pathway; positive regulation of hormone secretion; and vasopressin secretion) for cerebellum and cortex. One month after the exposure, both Gsc (Goosecoid Homeobox, connected with nervous system development) and Txlna (Taxilin Alpha, predicted to enable syntaxin binding activity) presented an increasing trend in dose for the hippocampus but decreased for the cerebellum. Gsc also showed a constant decrease in gene expression after six months of irradiation for the cortex. Tcof1 (Treacle Ribosome Biogenesis Factor 1, encodes a nucleolar protein with a LIS1 homology domain; required for neural crest specification) gene was present in every group independently of time and tissue, besides cerebellum at one month after the exposure. The Venn diagram in Figures 5A and 5B presents the observed similarity in dose trends.

Another dose-response pattern that has been often observed showed either an increase of gene expression for low dose compared to the control dose and a decrease for high dose compared to the control dose, or the opposite - a decrease of gene expression for low dose compared to control dose and increase for high dose compared to control dose (Figure 4, panels C and D). Such a pattern, which we named ∩-shaped and U-shaped, with different responses to corresponding doses, was also observed in each tissue (see Figure 5C and 5D). At one month after the exposure to irradiation, the ∩-shaped pattern was observed in 63 genes for the cerebellum, 50 genes for the cortex and 44 genes for the hippocampus, while the U-shaped pattern was present in 41 genes for the cerebellum, 58 genes for the cortex and 61 genes for the hippocampus (see Table 1). However, after six months, the ∩-shaped pattern was present in 36 genes for both the cerebellum and cortex and 40 genes for the hippocampus, while the U-shaped pattern was observed in 35 genes for the cerebellum, 62 genes for the cortex and 56 genes for the hippocampus (see Table 2). The full list of genes for each group can be seen in Tables S1 and S2. Repeatedly, the literature on the resulting genes was reviewed to find those associated with radiation-induced neurological functions. After one month of irradiation, the Ω-shaped pattern was present in Ecrg4 and Vmn1r43 (Vomeronasal type-1 receptor 43; putative pheromone receptor implicated in the regulation of social and reproductive behaviour) for both cortex and hippocampus, but Vmn1r43 also for cerebellum. Chrdl1 (Chordin Like 1, alters the fate commitment of neural stem cells from gliogenesis to neurogenesis) presented an example of such a trend in the cortex independently of time. After one month of irradiation, a few more genes were also presenting the Ω-shaped pattern, such as Npy6r (encodes Neuropeptide Y Receptor Y6, predicted to be involved in neuropeptide signalling pathway) for the cerebellum, Txlna and Gsc for cortex and Tcf4 for the hippocampus. That trend was also visible after six months of the exposure in Gsc for cerebellum, Hsd17b10 (encoding 3-hydroxyacyl-CoA dehydrogenase type II, a member of the short-chain dehydrogenase /reductase superfamily) for cortex and Txlna for hippocampus. The U-shaped pattern after one month of irradiation was present in Ecrg4 for the cerebellum, Tcf4 and Tti2 for the cortex and in Hsd17b10 for both the cortex and hippocampus. After six months of exposure to irradiation, that trend in dose was observed in Tti2 for the cerebellum, Ecrg4 and Gsc for the hippocampus, in Npy6r for both cerebellum and hippocampus and in Txlna for the cerebellum and cortex.

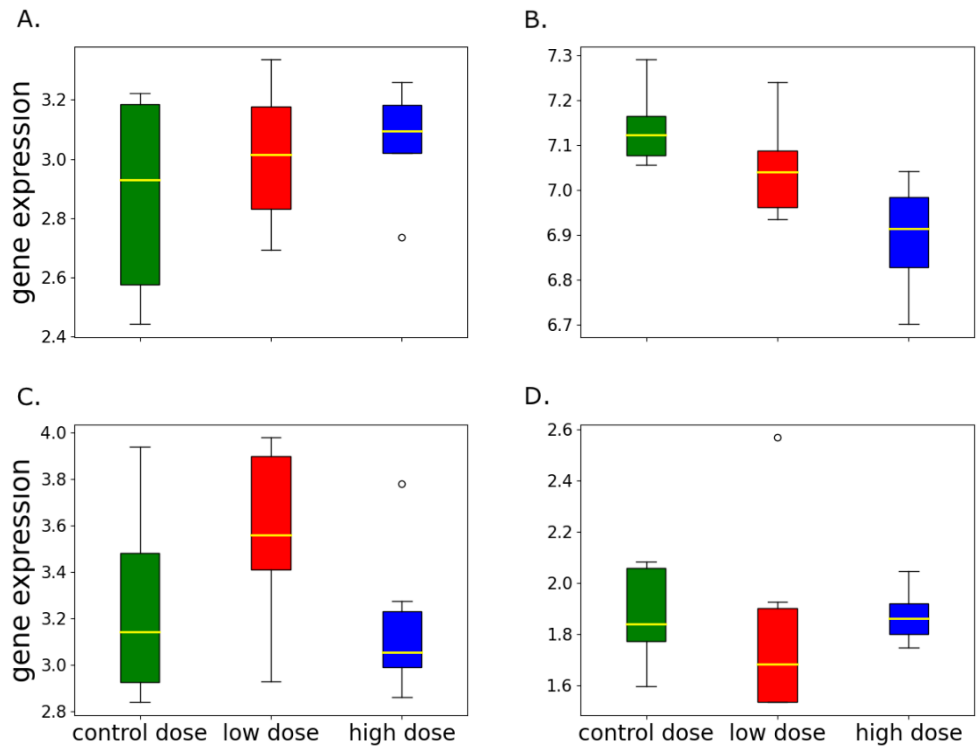

*Figure 4. Examples of different patterns in dose-response: additive increase of gene expression in the dose (panel A), additive decrease of gene expression within the dose (panel B), Ω-shaped (panel C) and U-shaped (panel D) dose responses.*

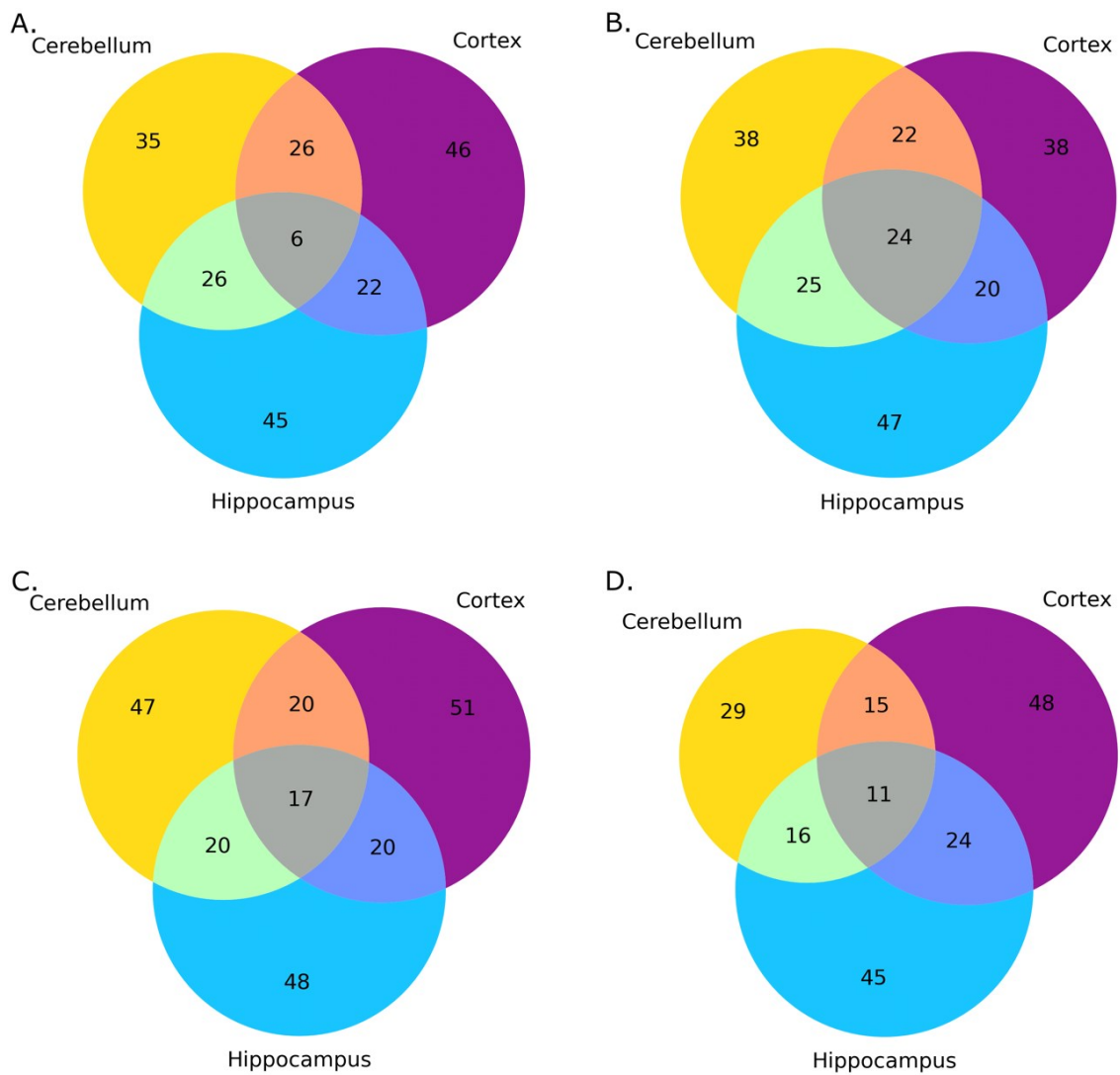

*Figure 5. Venn diagram of the number of genes in which either constant increase or decrease of the gene expression in dose has been observed after one month of irradiation (A.) and six months after irradiation (B.) or U-shape increase or decrease after one month of irradiation (C.) and six months after irradiation (D.).*

*Table 1. A number of genes in the studied brain tissues showing certain dose trends after one month of irradiation.*

|  | **Additive increase** | **Additive decrease** | **U-shaped** | **Ո-shaped** |
|---|---|---|---|---|
| **cerebellum** | 52 | 41 | 41 | 63 |
| **cortex** | 54 | 46 | 58 | 50 |
| **hippocampus** | 50 | 49 | 61 | 44 |

*Table 2. A number of genes in the studied brain tissues showing certain dose trends after six months of irradiation.*

|  | **Additive increase** | **Additive decrease** | **U-shaped** | **Ո-shaped** |
|---|---|---|---|---|
| **cerebellum** | 52 | 57 | 35 | 36 |
| **cortex** | 51 | 53 | 62 | 36 |
| **hippocampus** | 72 | 44 | 56 | 40 |

The study employed Gene Set Enrichment Analysis (GSEA) to conduct functional analysis six times for each time/tissue endpoint, which allowed comparison of the dose-dependent enrichment of gene sets with keywords related to the study's objective and to observe tissue differentiation in mice of different ages. A one-way ANOVA was performed for each specific brain tissue sample during the selected time interval, with dose as the independent factor and gene expression as the response variable. Gene sets specific to radiation exposure were identified as significant through GSEA and manual curation in tissues at one month (Figure 6A) and six months (Figure 6B) after exposure.

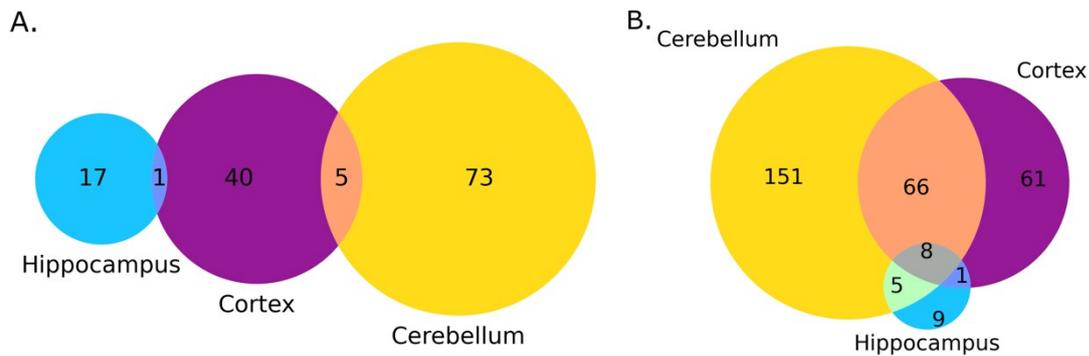

*Figure 6. A. Comparison of chosen significant gene sets after GSEA with radiation-specific gene sets, between tissues after one month of irradiation. B. Comparison of significant gene sets after GSEA with radiation-specific gene sets between tissues after six months of irradiation.*

The resulting pathways for every tissue and time interval after the irradiation can be studied in Tables S3-S8.

For each group that combines a specific tissue with a particular time period, the similarity coefficient was calculated for every pair of pathways in the significantly enriched gene sets. The similarity coefficient was determined as the fraction of genes shared by both pathways to the total number of unique genes present in those pathways. This approach allows for a comprehensive analysis of the pathways and their shared genes. The pathways were visualised using Cytoscape software. Nodes were used to represent the pathways and were connected by edges whose thickness was adjusted based on the similarity coefficient value. An edge threshold was applied to eliminate connections between gene sets that had no genes in common or had a very small similarity coefficient. This helped to improve the clarity and legibility of the visualisations. For all the tested tissues after 6 months of irradiation, the similarity coefficient threshold was set to 0.04. However, after 1 month of irradiation, the threshold for the cerebellum and cortex was set to 0.02, while for the hippocampus, it was set to 0.01 due to the limited number of significant pathways associated with the irradiation term. The networks for each tissue examined after 1 month of irradiation are presented in the left column of Figure 7, and after 6 months of irradiation, they are presented in the right column. Clusters related to changes in behaviour or the radiation response are visible in every network, although they are most prominent and dense in tissues 6 months after irradiation.

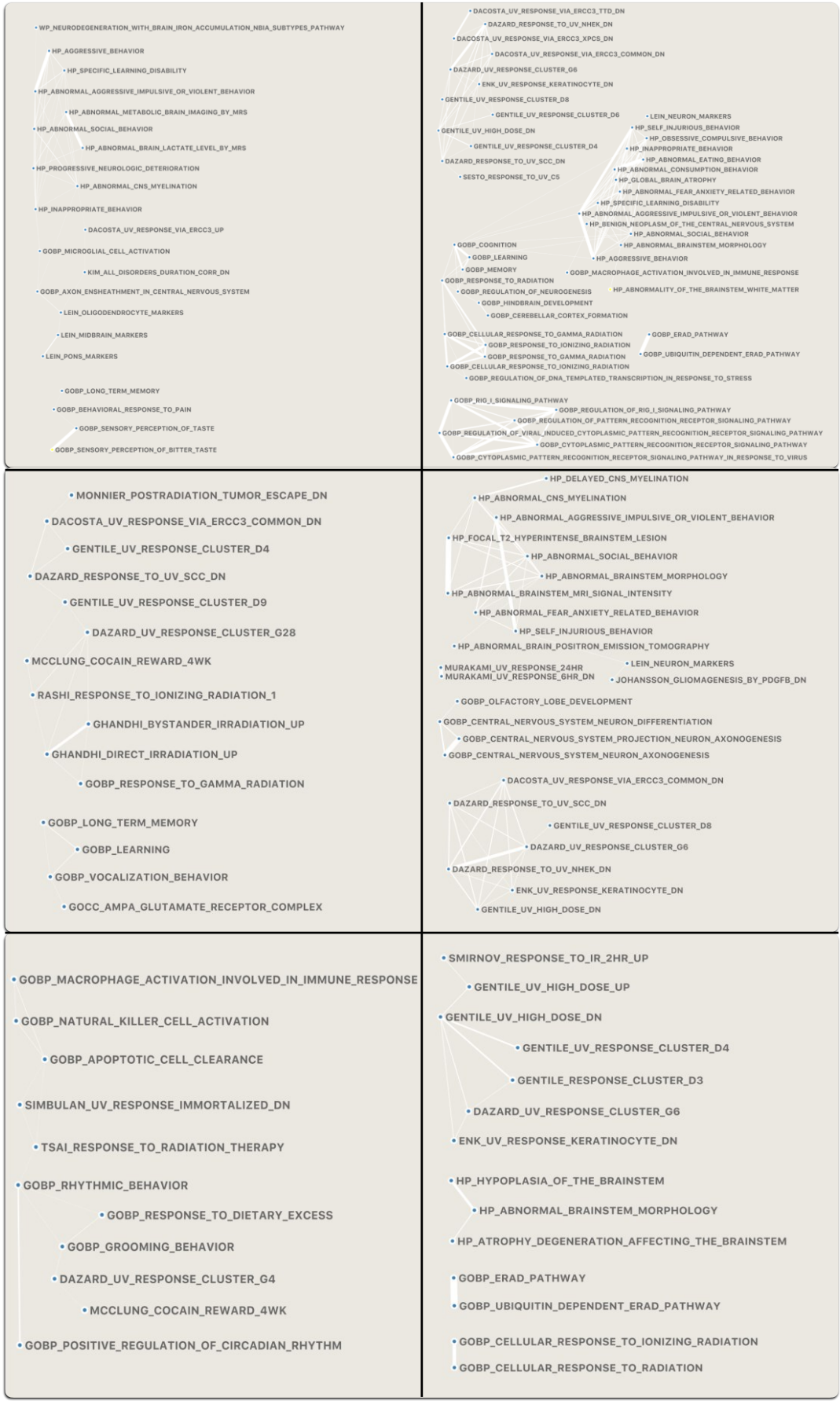

*Figure 7. Network presenting pathway connections for cerebellum (first row), cortex (second row) and hippocampus (third row) after one month of irradiation (left column) and six months after the irradiation (right column).*

# DISCUSSION

Since its discovery, ionising radiation has proven to be a double-edged sword. Its use in the medical field has meant a tremendous step forward in health care. However, it must be utilised with caution since exposure to radiation might as well provoke serious health consequences. In the first instance, the atomic bomb survivor studies have been indispensable in the evaluation of radiation-induced sequelae because of the use of large cohorts and a long-term follow-up. In these studies, radiation exposure was shown to induce cancer and leukaemia in a linear dose-dependent manner, although evidence for this becomes more ambiguous around doses below 100 mGy. Non-cancer effects, including cardiovascular diseases, cataracts, and postnatal/adult radiation-induced cognitive dysfunction, were also uncovered in epidemiological research but were believed to only arise after high-dose exposure. Recently, evidence accumulated that these non-cancer effects might as well be caused by doses around and below 0.5 Gy. As a general standard, pregnant women are better protected against ionising radiation than the general population because of the known adverse outcome for the unborn child, which has a high sensitivity to radiation-induced cancer. Importantly, besides cancer, prenatal irradiation can perturb brain development and cause neurological deficits, which were also demonstrated and evidenced for the first time by the atomic bomb survivor studies. On the other hand, we convincingly have shown in previous work in utero exposure to radiation-induced effects at low (0.1 Gy) and high (1.0 Gy) doses. In the short-term, 1.0 Gy irradiation (to a lesser extent 0.1 Gy) at E11 significantly induced DNA damage, cell cycle arrest and apoptosis, as well as an altered distribution of recently differentiated cells [1], [2], [3]. Mice exposed to 1.0 Gy (to a lesser extent 0.1 Gy) were defective in emotional responding and spatio-cognitive performance at a young adult age.

This paper attempts to illustrate the transcriptomic changes at 1 month and 6 months after birth in the brains of animals prenatally exposed to radiation. For that, combining the work of bioinformaticians and biologists was imperative. Advanced bioinformatics tools have been used for functional analysis. Still, the opinion and vision based on a biologist's background are highly appreciated in relation to pathways to answer the question in line with the project objectives. This ensures that the results are coherent and that the progress of the project can move in the desired direction.

The statistical analysis aimed to uncover the gene profile after irradiation in different brain regions at different time intervals after exposure to irradiation. The data was first visualised using dimensionality reduction techniques, which helped to provide an overview of the differentiation between factors. An analysis of variance (ANOVA) statistical test was applied to every gene to detect differences between the means of the experimental groups. The genes with at least a large effect size value only for the dose factor were taken into further analysis that had no interactions between the three differentiating factors. Various trends within doses were found and described.

Consecutively, functional analysis was performed, which sought to provide biological context to the groups that combined selected tissues with specific time intervals. A Gene Set Enrichment Analysis (GSEA) was performed for each group using gene sets related to irradiation to enrich the pathways. Statistically significant pathways were then manually curated by a biologist to select only those that fit the purpose of the study. Their networks with connections between pathways according to the similarity coefficient were visualised for each group. The formed clusters were then analysed, and each group that combined a certain tissue with a specific period of time was compared, to see how the tissues vary at different ages of mice. Of note, irradiation-induced apoptosis during gestation is considered protective to the developing embryo to eliminate damaged cells in a p53-dependent way [19], which might hamper further normal development. Therefore, we analysed the gene expression profiles in postnatal animals at 1 month and 6 months to establish a causal link between short- and long-term prenatal radiation effects. The identified GSEA and network pathways of the analysed brain regions (hippocampus, cortex and cerebellum) provided valuable information on brain behaviour-related pathways which are fully in line with previously performed behavioural testing and magnetic resonance imaging data [1], [2], [3]. For instance, in the cerebellum at 1 month, the most significant pathways are related to aggressive behaviour, abnormal social behaviour, learning disability, long-term memory, CNS myelination, oligodendrocyte and midbrain markers. At 6 months, these pathways appear less remarkable but persist with less significance for learning, aggressive behaviour, inappropriate behaviour and myelination. In the cortex, at 1 month, we mostly observe pathways involved in radiation (UV or Gamma) response, learning, long-term memory, and neuron development. In the 6 months, more significance is observed for pathways such as abnormal social behaviour, CNS myelination, neuron markers, aggressive, impulsive behaviour, brainstem markers, anxiety, axonogenesis and neuron differentiation. On the contrary, in the hippocampus, very few behaviour-related pathways are significant; at 1 month, we observe grooming behaviour, rhythmic behaviour and positive regulation of circadian rhythm; at 6 months, mainly pathways involved in hypoplasia and degeneration of the brainstem and abnormal brainstem development and morphology, indicating a defect in brain connections and brain communication. In all, we believe that the performed transcriptomic functional analysis strongly supports our previous observations by [1], [2], [3]. The identified pathways can be considered a valid biomarker for prenatal radiation-induced late cognitive impairment.

## CONCLUSIONS

It is extremely important to assess the impact of radiation on living organisms. Raising awareness of radiation defects in the developing brain is of utmost importance. This paper aims to illustrate the transcriptomic changes in the brain of animals prenatally exposed to radiation at 1 and 6 months after birth. Communication was effectively maintained between the bioinformatician and the biologist throughout the project, and frequent discussions were held to ensure progress in the right direction. It is recommended that both statistical and functional analysis be carried out to understand the results fully. The statistical analysis suggests that tissue plays a primary role in distinguishing gene expression, and each brain region exhibits varying degrees of sensitivity. The gene profiles revealed distinct responses to irradiation in different tissues. Detailed descriptions of the statistically significant pathway networks were provided for each tissue at various time intervals after radiation. The pathways that were found to be most activated appear to be related to behavioural changes and the response to irradiation. The results are satisfactory and strongly support previous observations. Therefore, the identified pathways could be considered a valid biomarker for prenatal radiation-induced late cognitive impairment.

## ACKNOWLEDGEMENT

Project was partially financed by Belgian Royal Decree 24.06.2021, Project ID: 21. Most of the transcriptomic data discussed in this paper were obtained within the FP7 EU CEREBRAD project (GA 295552)

This work has been partially supported by the ENEN2plus project (HORIZON-EURATOM-2021-NRT-01-13 101061677) founded by the European Union.

# BIBLIOGRAPHY

[1] Verreet, T., Verslegers, M., Quintens, R., Baatout, S., and Benotmane, M. A., 2016, "Current Evidence for Developmental, Structural, and Functional Brain Defects Following Prenatal Radiation Exposure," Neural Plast., **2016**.

[2] Verreet, T., Quintens, R., Van Dam, D., Verslegers, M., Tanori, M., Casciati, A., Neefs, M., Leysen, L., Michaux, A., Janssen, A., D'Agostino, E., Velde, G. Vande, Baatout, S., Moons, L., Pazzaglia, S., Saran, A., Himmelreich, U., De Deyn, P. P., and Benotmane, M. A., 2015, "A Multidisciplinary Approach Unravels Early and Persistent Effects of X-Ray Exposure at the Onset of Prenatal Neurogenesis," J. Neurodev. Disord., **7**(1), pp. 1–21.

[3] Verreet, T., Rangarajan, J. R., Quintens, R., Verslegers, M., Lo, A. C., Govaerts, K., Neefs, M., Leysen, L., Baatout, S., Maes, F., Himmelreich, U., D'Hooge, R., Moons, L., and Benotmane, M. A., 2016, "Persistent Impact of in Utero Irradiation on Mouse Brain Structure and Function Characterized by MR Imaging and Behavioral Analysis," Front. Behav. Neurosci., **10**(MAY), pp. 1–18.

[4] Russell, D. W., Kahn, J. H., Spoth, R., & Altmaier, E. M., 1998, "Analyzing Data from Experimental Studies: A Latent Variable Structural Equation Modeling Approach," J. Couns. Psychol.

[5] Belli, M., and Indovina, L., 2020, "The Response of Living Organisms to Low Radiation Environment and Its Implications in Radiation Protection," Front. Public Heal., **8**(December), pp. 1–15.

[6] Tapio, D. H. S., 2016, "Exposure of Ionizing Radiation on the Mammalian Brain: Epidemiological Evidence, Pathological and Molecular Effects and Prevention Strategies," Mutat. Res.

[7] M. Hossain, P. U. D., 2001, "Effect of Irradiation at the Early Foetal Stage on Adult Brain Function of Mouse: Learning and Memory," Int. J. Radiat. Biol.

[8] Andrzej Sienkiewicz, Ana Maria da Costa Ferreira, Birgit Danner, C. P. S., 1999, "Dielectric Resonator-Based Flow and Stopped-Flow EPR with Rapid Field Scanning: A Methodology for Increasing Kinetic Information," J. Magn. Reson.

[9] Andrzej Sienkiewicz, Kunbin Qu, and C. P. S., 1994, "Dielectric Resonatorbased Stoppedflow Electron Paramagnetic Resonance," Rev. Sci. Instrum.

[10] Surendran HP, Narmadha MP, Kalavagunta S, Sasidharan A, D. D., 2022, "Preservation of Cognitive Function after Brain Irradiation," J. Oncol. Pharm. Pract.

[11] Marazziti, D., Baroni, S., Catena-Dell'Osso, M., Schiavi, E., Ceresoli, D., Conversano, C., Dell'Osso, L., and Picano, E., 2012, "Cognitive, Psychological and Psychiatric Effects of Ionizing Radiation Exposure," Curr. Med. Chem., **19**(12), pp. 1864–1869.

[12] Marazziti, D., Piccinni, A., Mucci, F., Baroni, S., Loganovsky, K., and Loganovskaja, T., 2016, "Ionizing Radiation: Brain Effects and Related Neuropsychiatric Manifestations," Probl. Radiatsiinoi Medytsyny ta Radiobiolohii, **2016**(21), pp. 64–90.

[13] McInnes, L., Healy, J., and Melville, J., 2018, "UMAP: Uniform Manifold Approximation and Projection for Dimension Reduction."

[14] Cohen, J., 1988, *Statistical Power Analysis for the Behavioral Sciences*.

[15] Subramanian, A., Tamayo, P., Mootha, V. K., Mukherjee, S., Ebert, B. L., Gillette, M. A., Paulovich, A., Pomeroy, S. L., Golub, T. R., Lander, E. S., and Mesirov, J. P., 2005, "Gene Set Enrichment Analysis: A Knowledge-Based Approach for Interpreting

Genome-Wide Expression Profiles," Proc. Natl. Acad. Sci. U. S. A., **102**(43), pp. 15545–15550.
[16] Mootha, V., Lindgren, C., Eriksson, K. et al, 2003, "PGC-1α-Responsive Genes Involved in Oxidative Phosphorylation Are Coordinately Downregulated in Human Diabetes," Nat Genet.
[17] Koopmans, F., van Nierop, P., Andres-Alonso, M., Byrnes, A., Cijsouw, T., Coba, M. P., Cornelisse, L. N., Farrell, R. J., Goldschmidt, H. L., Howrigan, D. P., Hussain, N. K., Imig, C., de Jong, A. P. H., Jung, H., Kohansalnodehi, M., Kramarz, B., Lipstein, N., Lovering, R. C., MacGillavry, H., Mariano, V., Mi, H., Ninov, M., Osumi-Sutherland, D., Pielot, R., Smalla, K.-H., Tang, H., Tashman, K., Toonen, R. F. G., Verpelli, C., Reig-Viader, R., Watanabe, K., van Weering, J., Achsel, T., Ashrafi, G., Asi, N., Brown, T. C., De Camilli, P., Feuermann, M., Foulger, R. E., Gaudet, P., Joglekar, A., Kanellopoulos, A., Malenka, R., Nicoll, R. A., Pulido, C., de Juan-Sanz, J., Sheng, M., Südhof, T. C., Tilgner, H. U., Bagni, C., Bayés, À., Biederer, T., Brose, N., Chua, J. J. E., Dieterich, D. C., Gundelfinger, E. D., Hoogenraad, C., Huganir, R. L., Jahn, R., Kaeser, P. S., Kim, E., Kreutz, M. R., McPherson, P. S., Neale, B. M., O'Connor, V., Posthuma, D., Ryan, T. A., Sala, C., Feng, G., Hyman, S. E., Thomas, P. D., Smit, A. B., and Verhage, M., 2019, "SynGO: An Evidence-Based, Expert-Curated Knowledge Base for the Synapse.," Neuron, **103**(2), pp. 217-234.e4.
[18] "GeneCards – the Human Gene Database" [Online]. Available: https://www.genecards.org.
[19] Quintens, R., Verreet, T., Janssen, A., Neefs, M., Leysen, L., Michaux, A., Verslegers, M., Samari, N., Pani, G., Verheyde, J., Baatout, S., and Benotmane, M. A., 2015, "Identification of Novel Radiation-Induced P53-Dependent Transcripts Extensively Regulated during Mouse Brain Development," Biol. Open, **4**(3), pp. 331–344.

# Diversity in the radiation-induced transcriptomic temporal response of mouse brain tissue regions

Karolina Kulis, Sarah Baatout, Kevin Tabury, Joanna Polanska, Mohammed Abderrafi Benotmane

## SUPPLEMENTARY MATERIALS

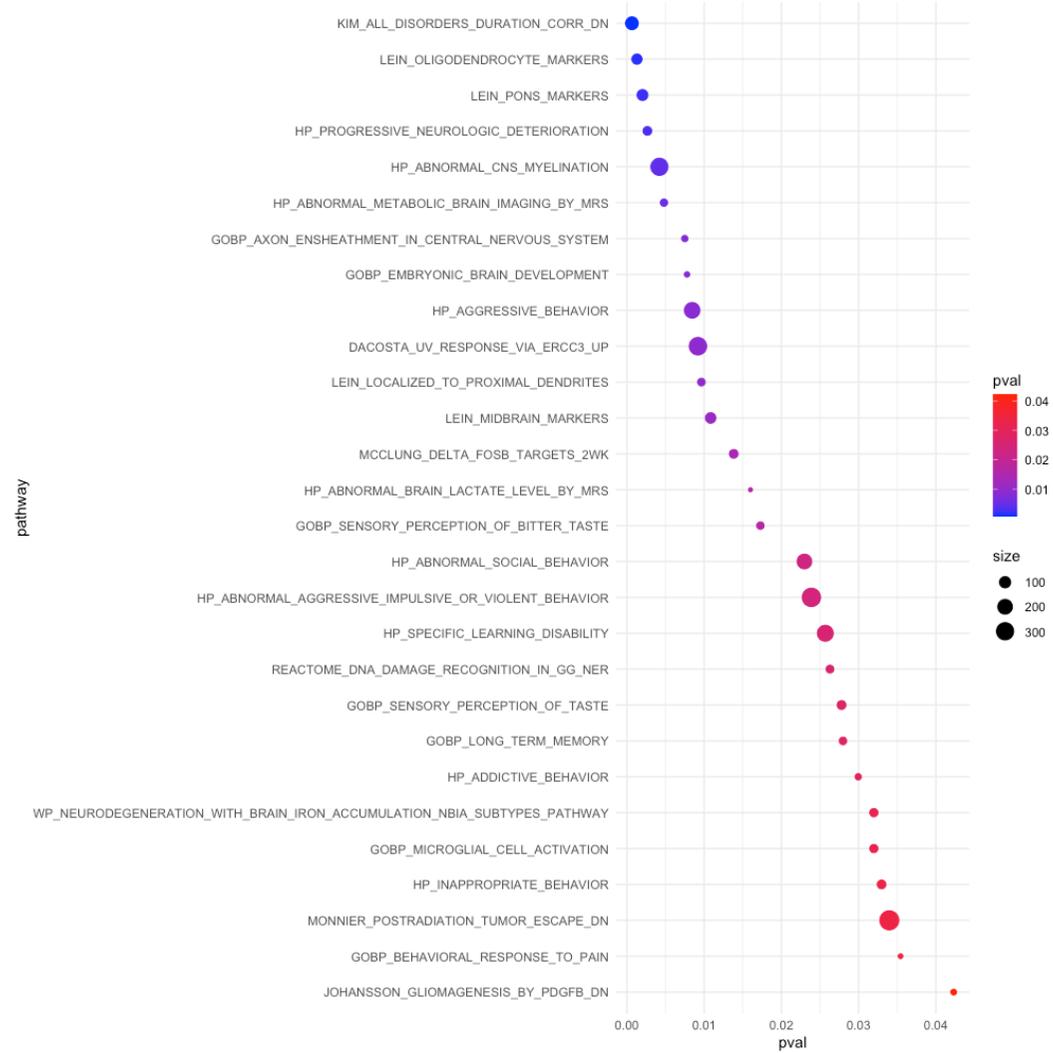

Figure S1. Dotplot presenting chosen significantly enriched pathways with p value < 0.05, for one month after irradiation for cerebellum

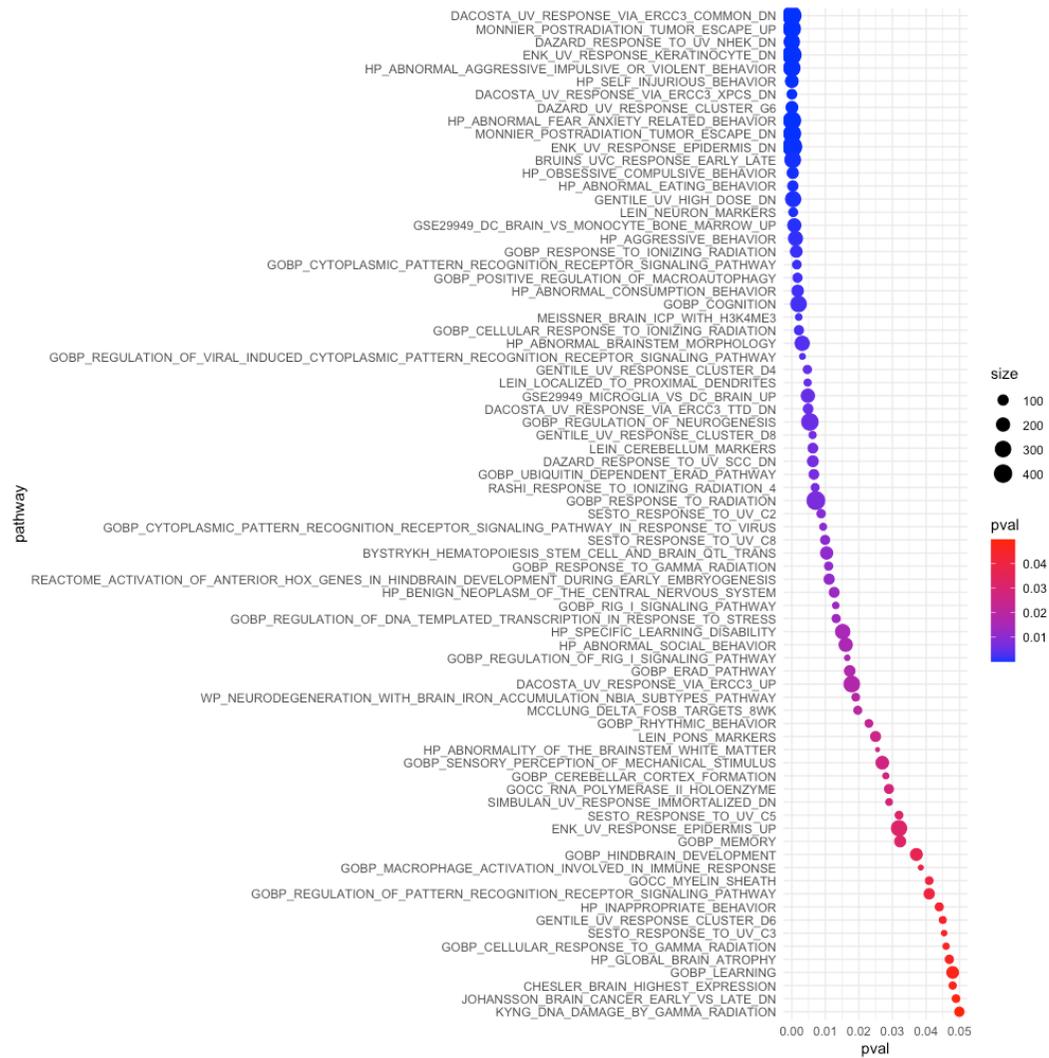

*Figure S2. Dotplot presenting chosen significantly enriched pathways with p value < 0.05, for six months after irradiation for cerebellum*

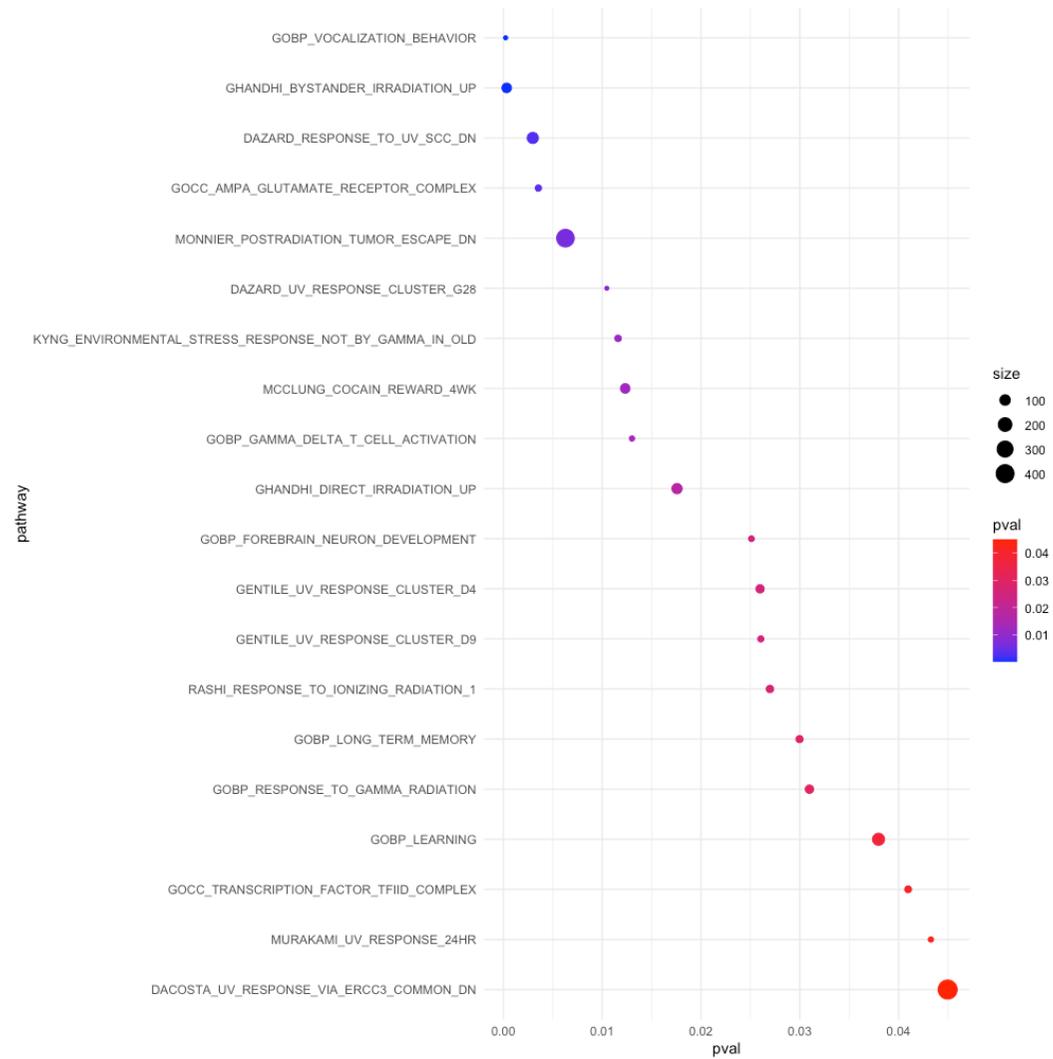

*Figure S3. Dotplot presenting chosen significantly enriched pathways with p value < 0.05, for one month after irradiation for cortex*

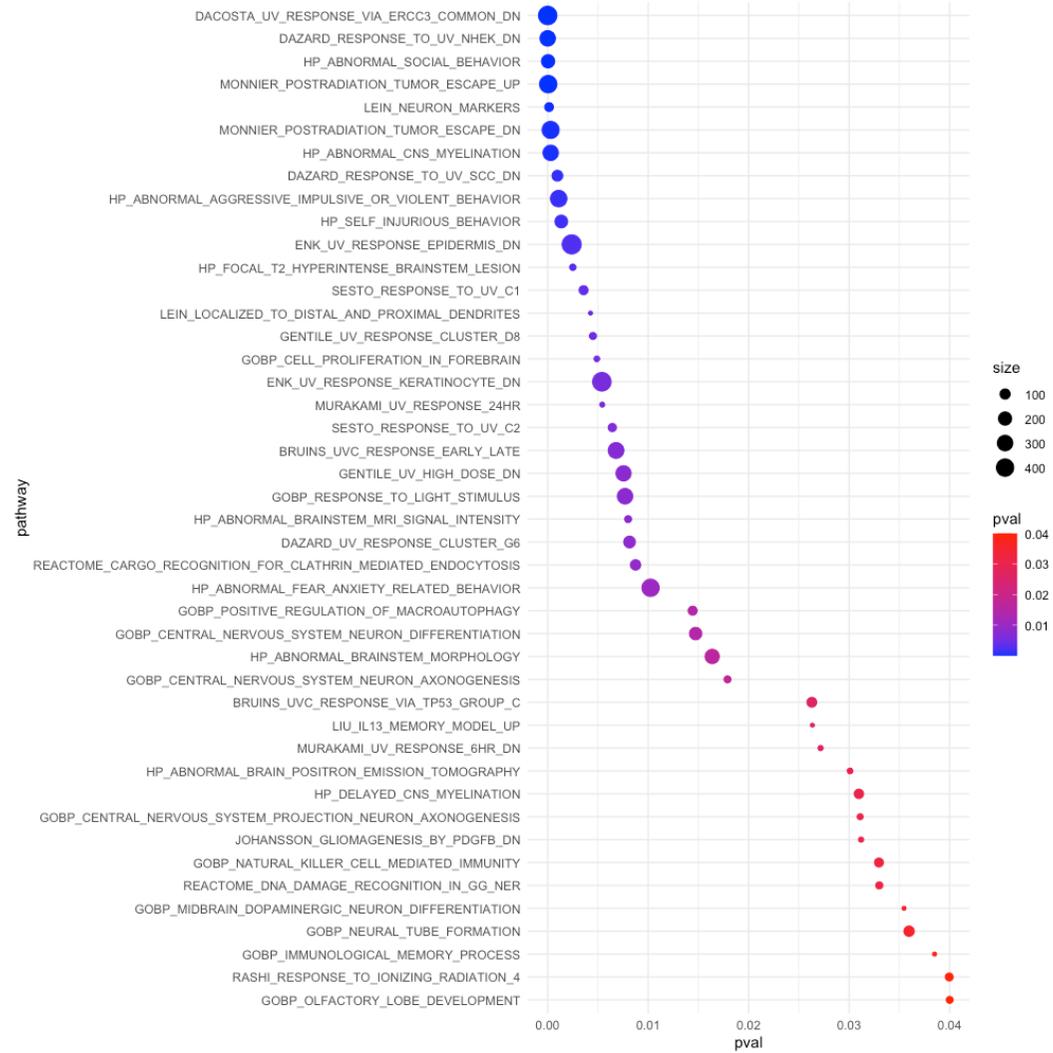

*Figure S4. Dotplot presenting chosen significantly enriched pathways with p value < 0.05, for six months after irradiation for cortex*

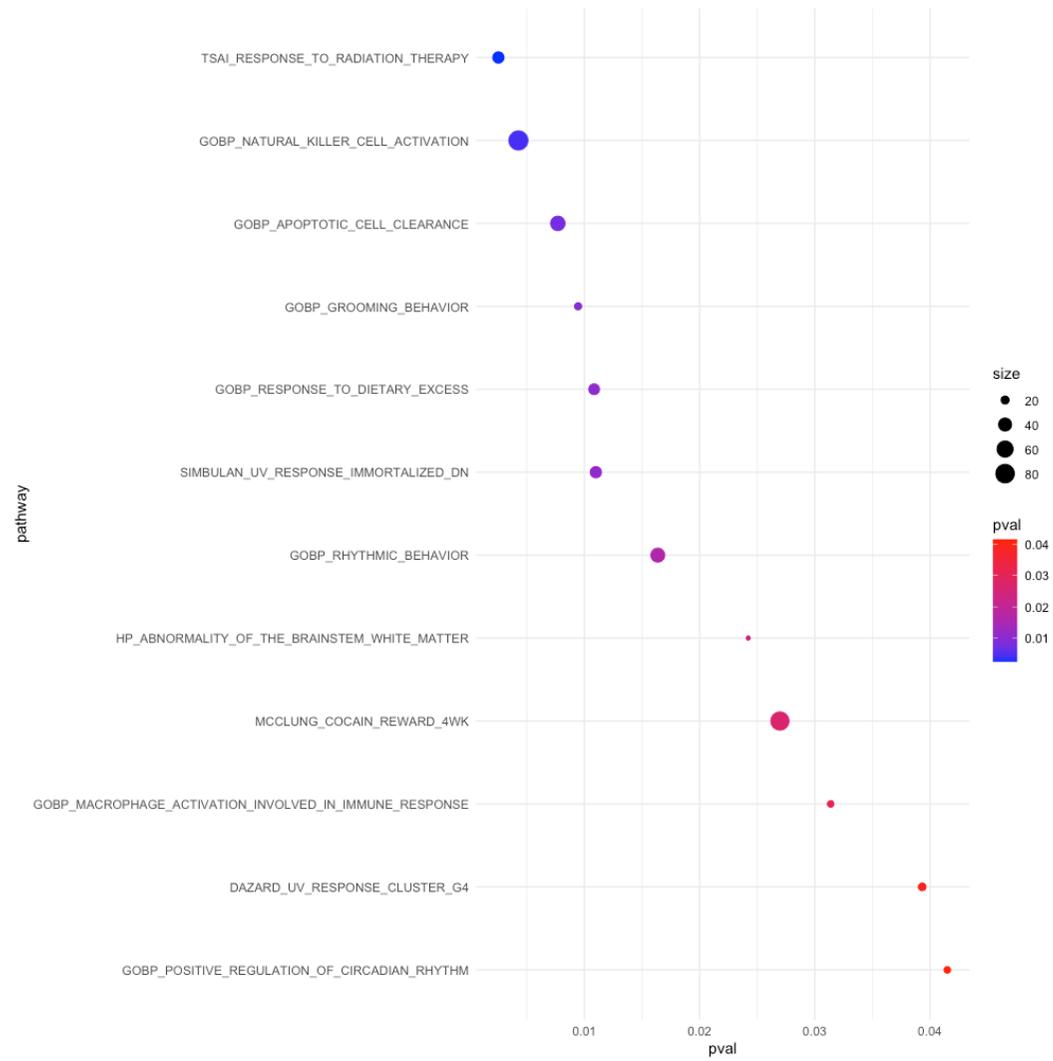

*Figure S5. Dotplot presenting chosen significantly enriched pathways with p value < 0.05, for one month after irradiation for hippocampus*

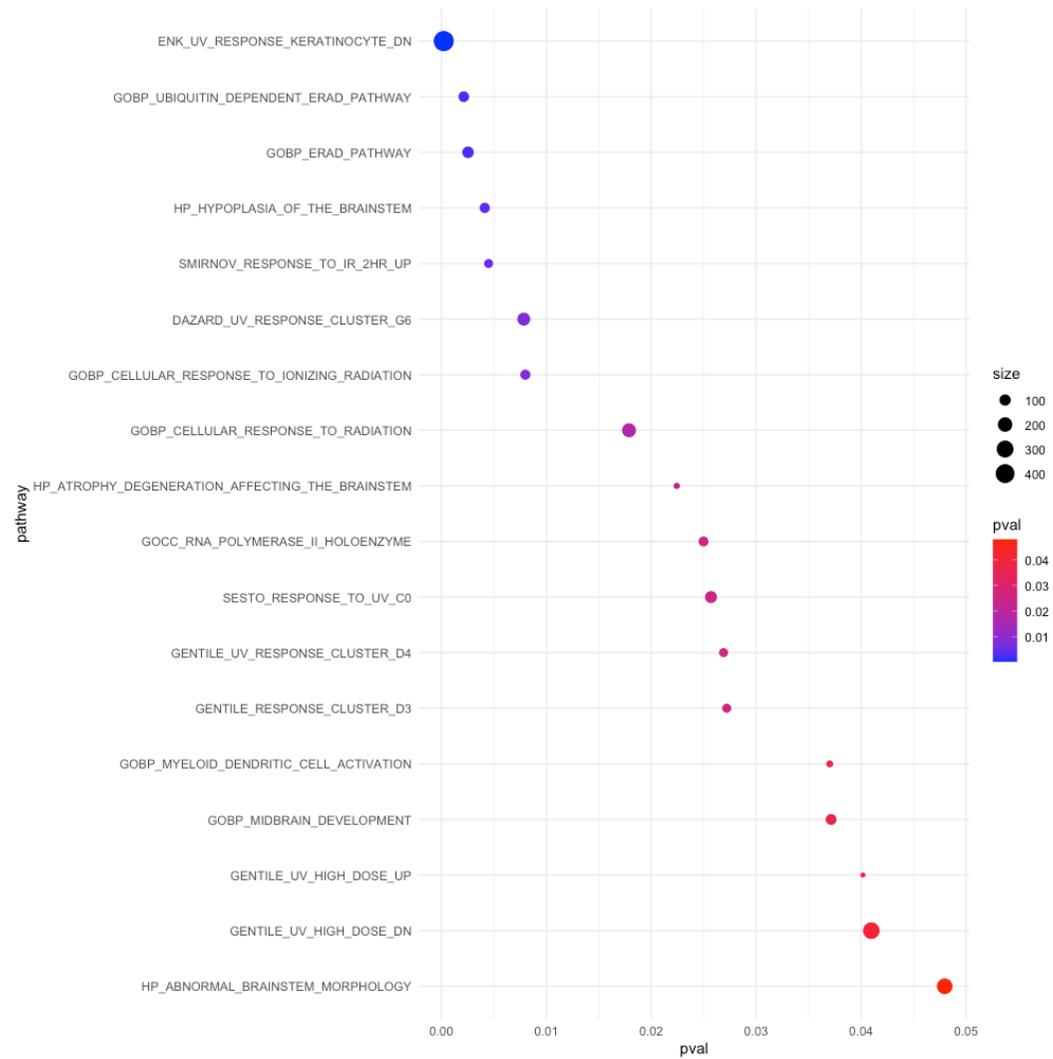

*Figure S6. Dotplot presenting chosen significantly enriched pathways with p value < 0.05, for six months after irradiation for hippocampus*

Table S1 List of genes with a particular trend. for a specific brain region. after one month of exposure to radiation.

| ascendingGenesCE1m | ascendingGenesCO1m | ascendingGenesHIP1m | descendingGenesCE1m | descendingGenesCO1m | descendingGenesHIP1m | UshapeCE1m | UshapeCO1m | UshapeHIP1m | ∩shapeCE1m | ∩shapeCO1m | ∩shapeHIP1m |
|---|---|---|---|---|---|---|---|---|---|---|---|
| 1700003D09Rik | 1700055D18Rik | 1700025F24Rik | 1700123M08Rik | 1700123M08Rik | 2810047C21Rik1 | 1700055D18Rik | 1700016L21Rik | 1700016L21Rik | 1700016L21Rik | 1700028P15Rik | 2310002F09Rik |
| A930033H14Rik | 4930448C13Rik | A930012L18Rik | 2310002F09Rik | 2810047C21Rik1 | Aard | 1700080G11Rik | 1700025F24Rik | 1700028P15Rik | 1700025F24Rik | 1700052K11Rik | AY512915 |
| Adamts2 | 4930548K13Rik | A930033H14Rik | AY512915 | A930012L18Rik | Aif1 | 4933428G20Rik | 1700055D18Rik | 1700055D18Rik | 1700028P15Rik | Aif1 | Apold1 |
| Akap3 | Accs | Accs | Aard | AY512915 | Brme1 | A930012L18Rik | 2310002F09Rik | 1700028P15Rik | 1700080G11Rik | Alg13 | Ascc3 |
| Alg13 | Adamts2 | Akap3 | Ccl11 | Aard | Ccl22 | Arfrp1 | 4930524J08Rik | 1700123M08Rik | 1700052K11Rik | Asb17 | C630031E19Rik |
| Avpi1 | BC064078 | Asb8 | Crnde | Apold1 | Crnde | BE692007 | 4933405L10Rik | 3110004A20Rik | 4930548K13Rik | Asb8 | Commd10 |
| BC052688 | Ccl11 | Aunip | Eif1ad | Arfrp1 | Gm10731 | Brme1 | 4933428G20Rik | 4930524J08Rik | 4933405L10Rik | Ascc3 | D1Pas1 |
| Bend3 | Ces2f | BC052688 | Glyr1 | Aunip | Gm11266 | Cfap36 | A930033H14Rik | Abcg2 | Asb17 | Ceacam-ps1 | Ecrg4 |
| Ccl22 | Eif1ad | Cfap36 | Gm10731 | Avpi1 | Gm11267 | Commd10 | Ago4 | Ago4 | Ctdp1 | Chrdl1 | Eef1akmt1 |
| Cdk11b | Foxn2 | Ctdp1 | Gm10791 | Cdk11b | Gm15587 | Ecrg4 | BE692007 | Arfrp1 | Cyp26a1 | Ecrg4 | Foxn2 |
| Cyp2d26 | Garin2 | Cyp2d26 | Gm11267 | Dhdds | Gm15589 | Eftud2 | Bend3 | Asb17 | Dhdds | Eef1akmt1 | Gm22187 |
| Fam163b | Gm10791 | Dhdds | Gm11954 | Esf1 | Gm16223 | Foxn2 | Brme1 | Avpi1 | Gm14936 | Eftud2 | Gm22345 |
| Gm13398 | Gm13398 | Dusp15 | Gm20110 | F630040K05Rik | Gm16279 | Fth1 | Ccdc8 | BE692007 | Gm15587 | Gm10126 | Gm24043 |
| Gm13974 | Gm1604a | Eftud2 | Gm22296 | Fam163b | Gm16299 | Gm16223 | Commd10 | CN725425 | Gm16299 | Gm10731 | Gm24285 |
| Gm14167 | Gm16279 | Esf1 | Gm23331 | Fth1 | Gm20110 | Gm16279 | Crnde | Ccl11 | Gm17249 | Gm15589 | Gm25150 |
| Gm16364 | Gm20110 | Fam163b | Gm24285 | Gm13974 | Gm22127 | Gm20751 | Cdk11b | Ces2f | Gm20026 | Gm16982 | Gm25152 |
| Gm22004 | Gm22004 | Fyb1 | Gm25622 | Gm14936 | Gm22544 | Gm23460 | Dusp15 | Defb40 | Gm22127 | Gm20026 | Gm25811 |
| Gm22544 | Gm22127 | Gm20026 | Gm25919 | Gm17249 | Gm24949 | Gm24523 | Gm14167 | Egfl7 | Gm22498 | Gm24285 | Gm26437 |
| Gm23709 | Gm22498 | Gm22296 | Gm26437 | Gm22915 | Gm25058 | Gm24949 | Gm15587 | Eif1ad | Gm23453 | Gm25058 | Gm50619 |
| Gm24230 | Gm22296 | Gm23453 | Gm57518 | Gm23709 | Gm25431 | Hmgb1 | Gm16223 | Fth1 | Gm23895 | Gm25269 | Gm5451 |
| Gm25152 | Gm23331 | Gm23453 | Gm8700 | Gm23895 | Gm25600 | Ighv1-74 | Gm20751 | Gm10791 | Gm24043 | Gm25431 | Gm6605 |
| Gm25431 | Gm23453 | Gm24230 | Gm9234 | Gm25150 | Gm26367 | Inka2 | Gm22281 | Gm22296 | Gm24481 | Gm25622 | Gm8700 |
| Gm25811 | Gm24523 | Gm2710 | Gsc | Gm26437 | Herc6 | Lypd8 | Gm22915 | Gm10791 | Gm24840 | Gm11954 | Ift46 |
| Gm26367 | Gm25150 | Gm38388 | Lin54 | Gm50619 | Igfbp7 | Mir25 | Gm25152 | Gm11954 | Gm25269 | Gm25600 | Igkv4-55 |
| Gm38388 | Gm25811 | Gsc | Or10n1 | Gm5930 | Ighv1-18 | Lypd8 | Gm14936 | Gm26411 | Gm47655 | Gm9767 | Mir25 |
| Gm6605 | Gm26367 | H2-Ab1 | Or10p21 | Hacd4 | Ighv1-74 | Myl4 | Gm25802 | Gm1604a | Gm25600 | Gsc | Mir5135 |
| Lrrc32 | Gm57518 | Lin54 | Or52a5b | Herc6 | Ighv9-3 | Naa20 | Gm25924 | Gm16364 | Gm5766 | Igfbp7 | Mrps14 |
| Mark4 | Gm5766 | Lypd8 | Or8b55 | Inka2 | Mark4 | Or11i1 | Gm6605 | Gm17249 | Gm7160 | Igkv4-55 | Myh13 |
| Mccc1 | Gm7160 | Mad2l1bp | Rpl39l | Mccc1 | Meis1 | Or2ak5 | Gm17249 | Gm20751 | Gm9767 | Krtap19-1 | Or10p21 |
| Mir204 | Gsta2 | Myl4 | Stard8 | Mill2 | Mir1187 | Or5p81 | Gm22915 | Gm7337 | H2-Ab1 | Mrps14 | Or51a5 |
| Mir5100 | Hmgb1 | Neo1 | Sycp1-ps1 | Myh13 | Mir93 | Or7g19 | Hsd17b10 | Gm23331 | Hacd4 | Mrps14 | Nr5a2 |
| Mir763 | Ighv9-3 | Nlrp4c | Tdpoz3 | Myl4 | Myom1 | Or2ak5 | Ift46 | Gm23460 | Hnf4g | Or10ak11 | Or52ab2 |
| Muc20 | Krtap11-1 | Oga | Tex264 | Myom1 | Naa20 | Plxna4os1 | Ighv1-18 | Gm25802 | Ighv9-3 | Or13l2 | Or5p81 |
| Myh13 | Mir5100 | Or4c1 | Traj61 | Nodal | Nodal | Ptrh2 | Ighv1-18 | Gm25919 | Kif18b | Or6c3b | Or6c3b |
| Ncapd3 | Mir763 | Or8k38 | Txlna | Omp | Npy6r | Rpia | Kif18b | Gm26411 | Mad2l1bp | Or52a24 | Or8b55 |
| Nup210l | Mir767 | Phf6 | Utp14b | Or1l8 | Nr5a2 | Sf3a3 | Lypd8 | Gm5930 | Mccc2 | Pds5a | Parm1 |
| Oga | Muc20 | Pradc1 | Vcf1 | Or51a5 | Or10n1 | Spata25 | Mark4 | Gsta2 | Mill2 | Phf6 | Pds5a |
| Or13l2 | Nodal | Rhox4c | Wdr5b | Or8k38 | Or1l8 | Tcof1 | Mccc2 | Hsd17b10 | Mir1954 | Plscr1l1 | Plscr1l1 |
| Or51a5 | Nup210l | Rpia | Zfp1001 | Pantr1 | Or2d2b | Thoc2 | Mir1187 | Mir1954 | Mir767 | Prr23a2 | Snord22 |
| Or52a24 | Oga | Spata25 | Zfp422-ps | Peg13 | Or52a24 | Thsd1 | Mir93 | Mir204 | Mir93 | Prss52 | Tank |
| Or8k3 | Or10p21 | Stard8 | n-R5s5 | Rassf4 | Or5ac17 | Tmem171 | Ncapd3 | Mrps36-ps1 | Mrps14 | Stard8 | Tcf4 |
| Oxct1as | Or2z8 | Sycp1-ps1 | | Sec11c | Or7g19 | Trmt6 | Neo1 | Mrps36-ps1 | Mrps14 | Stard8 | Tmem267 |
| Peg13 | Or52ab2 | Tdpoz3 | | Snord22 | Pantr1 | | Nlrp4c | Ncapd3 | Nlrp4c | Sycp1-ps1 | Vmn1r43 |
| Rufy4 | Or8b55 | Tmem171 | | Tank | Plxna4os1 | | Or10n1 | Nup210l | Npy6r | Tm4sf4 | Vmn2r109 |
| Spata31g1 | Otub1 | Tmprss9 | | Tcof1 | Psma4 | | Or2ak5 | Or10ak11 | Or2ak5 | Txlna | Vmn2r77 |
| Srsf5 | Parm1 | Tti2 | | Vmn2r86 | Rpl39l | | Or4c1 | Or2d2b | Or4c1 | Utp14b | |
| Tas2r107 | Rpl39l | Txlna | | Zfp276 | Tcl1b2 | | Or5ac17 | Or52a5b | Or52ab2 | Vmn1r43 | |
| Tcf4 | Spata25 | Usp42 | | | Tcof1 | | Or5p81 | Otub1 | Or6c3b | Vmn2r109 | |
| Tcl1b2 | Spinkl | Vmn1r32 | | | Tm4sf4 | | Pglyrp4 | Peg13 | Otub1 | Zfp1001 | |
| Tm4sf4 | Tafazzin | Vmn2r15 | | | | | Psma4 | Pglyrp4 | Pantr1 | Zfp422-ps | |
| Tti2 | Tmem171 | Zfp1001 | | | | | Spata31g1 | Pltp | Parm1 | | |
| n-R5s65 | Ttll8 | | | | | | Srsf5 | Prr23a2 | Pds5a | | |
| | Usp42 | | | | | | Tcf4 | Rprd2 | Pglyrp4 | | |
| | Vmn1r32 | | | | | | Tex264 | Sf3a3 | Poteg | | |
| | Vmn2r77 | | | | | | Thsd1 | Snora20 | Prrt1b | | |
| | | | | | | | Tmprss9 | Srsf5 | Ptchd4 | | |
| | | | | | | | Traj61 | Tas2r107 | Rhox4c | | |
| | | | | | | | Trmt6 | Trim71 | Snord22 | | |
| | | | | | | | Tti2 | Trmt6 | Tank | | |
| | | | | | | | Vcf1 | Vcf1 | Tmprss9 | | |
| | | | | | | | | Zfp276 | Usp42 | | |
| | | | | | | | | Zfp422-ps | Vmn1r32 | | |
| | | | | | | | | n-R5s65 | Vmn1r43 | | |
| | | | | | | | | | Vmn2r109 | | |

*Table S2 List of genes with a particular trend. for a specific brain region. after six months of exposure to radiation.*

| ascendingGenesCE6m | ascendingGenesCO6m | ascendingGenesHIP6m | descendingGenesCE6m | descendingGenesCO6m | descendingGenesHIP6m | UshapeCE6m | UshapeCO6m | UshapeHIP6m | ∩shapeCE6m | ∩shapeCO6m | ∩shapeHIP6m |
|---|---|---|---|---|---|---|---|---|---|---|---|
| 1700080G11Rik | A930033H14Rik | 1700016L21Rik | 1700016L21Rik | 1700123M08Rik | 4933405L10Rik | Arfrp1 | 1700016L21Rik | 1700055D18Rik | 1700055D18Rik | 1700003D09Rik | 1700080G11Rik |
| 1700123M08Rik | Aard | 2810047C21Rik1 | 4930448C13Rik | 2310002F09Rik | 4933428G20Rik | BE692007 | 1700025F24Rik | 1700028P15Rik | 2310002F09Rik | 1700028P15Rik | 2310002F09Rik |
| 4930524J08Rik | Ago4 | 4930448C13Rik | Aard | 3110004A20Rik | Aunip | Cyp26a1 | 1700052K11Rik | 1700052K11Rik | AY512915 | 2810047C21Rik1 | AY512915 |
| 4930548K13Rik | Arfrp1 | 4930548K13Rik | Asb17 | 4933405L10Rik | Dipk2a | Foxn2 | 1700055D18Rik | 1700055D18Rik | 3110004A20Rik | Aunip | Akap3 |
| 4933428G20Rik | Asb8 | A930033H14Rik | Aunip | 4933428G20Rik | E430024I08Rik | Gm16982 | Apold1 | 1700123M08Rik | Accs | BE692007 | CN725425 |
| A930012L18Rik | Ceacam23 | Accs | CN725425 | A930012L18Rik | Esf1 | Gm20026 | BC064078 | 4930430F21Rik | Ceacam23 | Cfap36 | Cdk11b |
| Adamts2 | D1Pas1 | Ago4 | Ccl11 | AY512915 | Fyb1 | Gm20110 | C630031E19Rik | 4930524J08Rik | Commd10 | Chrdl1 | Ces2f |
| Ago4 | Ecrg4 | Alg13 | Ccl22 | Abcg2 | Gm11954 | Gm20751 | CN725425 | A930012L18Rik | Crnde | Dipk2a | Cfap36 |
| Ctdp1 | Eftud2 | Avpi1 | Cfap36 | Accs | Gm13398 | Gm22544 | Crnde | Apold1 | Gm11954 | Eif1ad | Crnde |
| Cyp2d26 | Egfl7 | BE692007 | Chrdl1 | Asb17 | Gm15587 | Gm23453 | Cyp26a1 | Ceacam-ps1 | Gm1604a | F630040K05Rik | Eftud2 |
| Ecrg4 | Fth1 | Brme1 | D1Pas1 | BC052688 | Gm16364 | Gm24523 | Gm1604a | Ceacam23 | Gm17249 | Fam163b | Egfl7 |
| Eftud2 | Gm10731 | C630031E19Rik | Defb40 | Bend3 | Gm23460 | Gm25622 | Gm16279 | Dusp15 | Gm24230 | Fyb1 | Fth1 |
| Egfl7 | Gm14167 | Ccdc8 | Dipk2a | Ccdc8 | Gm24043 | Gm25811 | Gm16364 | Ecrg4 | Gm25269 | Glyr1 | Garin2 |
| Fam163b | Gm16223 | Ctdp1 | E430024I08Rik | Ctdp1 | Gm26367 | Gm25924 | Gm16982 | Eef1akmt1 | Gm25919 | Gm11266 | Gm11266 |
| Fyb1 | Gm22127 | Cyp26a1 | Gm10466 | Cyp2d26 | Gm26437 | Igfbp7 | Gm17249 | F630040K05Rik | Gm5451 | Gm13974 | Gm14936 |
| Gm13974 | Gm22296 | Dhdds | Gm10791 | Gm20026 | Gm38388 | Kif18b | Gm20110 | Foxn2 | Gm57518 | Gm25150 | Gm20026 |
| Gm14167 | Gm22345 | Eif1ad | Gm14936 | Gm22004 | H2-Ab1 | Mark4 | Gm20751 | Gm14167 | Gm6605 | Gm3134 | Gm22281 |
| Gm16279 | Gm22498 | Fam163b | Gm22127 | Gm24043 | Hacd4 | Mccc1 | Gm22187 | Gm15589 | Gm7160 | Gm8700 | Gm22544 |
| Gm22296 | Gm24285 | Gm11267 | Gm22187 | Gm24230 | Inka2 | Meis1 | Gm22281 | Gm16223 | Gm9234 | Gsta2 | Gm25058 |
| Gm22498 | Gm25600 | Gm13974 | Gm23460 | Gm24481 | Kif18b | Mill2 | Gm22915 | Gm16279 | Gsc | Herc6 | Gm25431 |
| Gm26411 | Gm25802 | Gm20110 | Gm23709 | Gm25622 | Krtap19-1 | Myom1 | Gm23331 | Gm16982 | Herc6 | Hsd17b10 | Gm2710 |
| Gm47655 | Gm25811 | Gm22127 | Gm24043 | Gm26411 | Lin54 | Npy6r | Gm23460 | Gm17249 | Hnf4g | Ighv1-18 | Gm5930 |
| Gm5766 | Gm25919 | Gm22296 | Gm24285 | Gm2710 | Mill2 | Nr5a2 | Gm23709 | Gm22345 | Mir1187 | Inka2 | Gm6605 |
| Ighv1-18 | Gm26437 | Gm22498 | Gm24949 | Gm47655 | Mir1187 | Or52a5b | Gm23895 | Gm23709 | Mir25 | Myh13 | Gm9767 |
| Inka2 | Ift46 | Gm22915 | Gm25150 | Gm50619 | Mir93 | Or7g19 | Gm24523 | Gm24230 | Myl4 | Myl4 | Herc6 |
| Lin54 | Igfbp7 | Gm23331 | Gm25431 | Gm7160 | Mrps14 | Prr23a2 | Gm24840 | Gm24285 | Or2ak5 | Oga | Hnf4g |
| Lrrc32 | Kif18b | Gm25600 | Gm25802 | Gm7337 | Muc20 | Rpia | Gm25058 | Gm25152 | Or8b55 | Or2ak5 | Ighv9-3 |
| Mir1954 | Lypd8 | Gm25802 | Gm26367 | Gm9767 | Nodal | Snord22 | Gm25269 | Gm25269 | Rassf4 | Or7g19 | Igkv4-55 |
| Mir763 | Mccc1 | Gm25811 | Gm2710 | Gsc | Oga | Srsf5 | Gm25431 | Gm50619 | Spata25 | Or8k3 | Lypd8 |
| Mir93 | Mccc2 | Gm3134 | Gm5930 | Igkv4-55 | Or4c1 | Thsd1 | Gm25924 | Gm5451 | Speer9-ps1 | Peg13 | Meis1 |
| Naa20 | Mill2 | Gm47655 | Gm7337 | Lin54 | Or8b55 | Tti2 | Gm38388 | Gsc | Stard8 | Rassf4 | Or10ak11 |
| Ncapd3 | Mir1187 | Gm7160 | Krtap19-1 | Lrrc32 | Pds5a | Txlna | Gm5930 | Ift46 | Tank | Rprd2 | Or10n1 |
| Nlrp4c | Naa20 | Gm7337 | Mir767 | Mark4 | Pglyrp4 | Usp42 | Hacd4 | Igfbp7 | Tmprss9 | Stard8 | Pantr1 |
| Omp | Nlrp4c | Gm8700 | Mrps14 | Mir25 | Poteg | Utp14b | Hnf4g | Ighv1-74 | Ttll8 | Traj61 | Rhox4c |
| Or10n1 | Or5p81 | Gm9234 | Mrps36-ps1 | Nodal | Prrt1b | Vmn2r77 | Ighv1-74 | Lrrc32 | Zfp422-ps | Vcf1 | Rpia |
| Or11i1 | Or8b55 | Hsd17b10 | Myh13 | Or11i1 | Rufy4 | | Mir204 | Mad2l1bp | n-R5s65 | Vmn2r109 | Sf3a3 |
| Or52ab2 | Otub1 | Ighv1-18 | Neo1 | Or13l2 | Sec11c | | Mir763 | Mir5135 | | | Snord22 |
| Oxct1as | Phf6 | Mccc1 | Nodal | Or2d2b | Spata31g1 | | Mir93 | Mir763 | | | Tm4sf4 |
| Plxna4os1 | Rpl39l | Mccc2 | Or10p21 | Or51a5 | Tcof1 | | Muc20 | Npy6r | | | Txlna |
| Prrt1b | Spinkl | Mir1954 | Or51a5 | Or52a24 | Tmem267 | | Ncapd3 | Nr5a2 | | | Vmn2r109 |
| Snora20 | Tafazzin | Mir204 | Or52a24 | Or52ab2 | Tmprss9 | | Omp | Or10p21 | | | |
| Spata31g1 | Tank | Mir5100 | Or6c3b | Or5ac17 | Zfp422-ps | | Or10ak11 | Or13l2 | | | |
| Spinkl | Tas2r107 | Mrps36-ps1 | Or8k38 | Prrt1b | n-R5s5 | | Or10n1 | Or2z8 | | | |
| Tafazzin | Tdpoz3 | Myh13 | Pglyrp4 | Rufy4 | n-R5s65 | | Or10p21 | Parm1 | | | |
| Tcf4 | Thoc2 | Myl4 | Plscr1l1 | Sf3a3 | | | Or1l8 | Plscr1l1 | | | |
| Tm4sf4 | Thsd1 | Myom1 | Poteg | Speer9-ps1 | | | Or2z8 | Pltp | | | |
| Trim71 | Tm4sf4 | Nlrp4c | Ptchd4 | Tcof1 | | | Or52a5b | Pradc1 | | | |
| Trmt6 | Trmt6 | Nup210l | Ptrh2 | Tmem267 | | | Or8k38 | Prr23a2 | | | |
| Vmn2r109 | Wdr5b | Or11i1 | Rufy4 | Utp14b | | | Oxct1as | Prss52 | | | |
| Vmn2r86 | Zfp1001 | Or2ak5 | Sf3a3 | Vmn1r43 | | | Pglyrp4 | Ptrh2 | | | |
| Wdr5b | Zfp422-ps | Or2d2b | Sycp1-ps1 | Vmn2r77 | | | Pltp | Rassf4 | | | |
| Zfp276 | | Or51a5 | Tcof1 | Vmn2r86 | | | Pradc1 | Rpl39l | | | |
| | | Or52a24 | Tdpoz3 | n-R5s5 | | | Ptchd4 | Tank | | | |
| | | Or52a5b | Tmem171 | | | | Rpia | Tex264 | | | |
| | | Or5ac17 | Tmem267 | | | | Snora20 | Vmn2r15 | | | |
| | | Or7g19 | Zfp1001 | | | | Snord22 | Zfp276 | | | |
| | | Or8k3 | n-R5s5 | | | | Srsf5 | | | | |
| | | Otub1 | | | | | Tcl1b2 | | | | |
| | | Oxct1as | | | | | Tex264 | | | | |
| | | Plxna4os1 | | | | | Ttll8 | | | | |
| | | Psma4 | | | | | Txlna | | | | |
| | | Ptchd4 | | | | | n-R5s65 | | | | |
| | | Rprd2 | | | | | | | | | |
| | | Speer9-ps1 | | | | | | | | | |
| | | Tas2r107 | | | | | | | | | |

S9

| ascendingGenesCE6m | ascendingGenesCO6m | ascendingGenesHIP6m | descendingGenesCE6m | descendingGenesCO6m | descendingGenesHIP6m | UshapeCE6m | UshapeCO6m | UshapeHIP6m | ∩shapeCE6m | ∩shapeCO6m | ∩shapeHIP6m |
|---|---|---|---|---|---|---|---|---|---|---|---|
| | | Tdpoz3 | | | | | | | | | |
| | | Traj61 | | | | | | | | | |
| | | Tti2 | | | | | | | | | |
| | | Usp42 | | | | | | | | | |
| | | Vcf1 | | | | | | | | | |
| | | Vmn2r86 | | | | | | | | | |
| | | Wdr5b | | | | | | | | | |

*Table S3 Selected radiation-specific significant gene sets after GSEA. for cerebellum after one month of radiation exposure.*

| pathway | pval | fdr | log2err | ES | NES | size |
|---|---|---|---|---|---|---|
| MONNIER_POSTRADIATION_TUMOR_ESCAPE_DN | 0.033966034 | 0.322177822 | 0.245041785 | 0.354919345 | 1.114185484 | 380 |
| HP_ABNORMAL_AGGRESSIVE_IMPULSIVE_OR_VIOLENT_BEHAVIOR | 0.023881168 | 0.322177822 | 0.352487858 | 0.360232821 | 1.128615482 | 348 |
| HP_SPECIFIC_LEARNING_DISABILITY | 0.025683497 | 0.322177822 | 0.352487858 | 0.373716324 | 1.158964514 | 246 |
| DACOSTA_UV_RESPONSE_VIA_ERCC3_UP | 0.009191707 | 0.270341171 | 0.380730401 | 0.375425976 | 1.172239781 | 306 |
| HP_ABNORMAL_CNS_MYELINATION | 0.004201468 | 0.225828894 | 0.407017919 | 0.380881784 | 1.186471926 | 293 |
| HP_AGGRESSIVE_BEHAVIOR | 0.00844443 | 0.270341171 | 0.380730401 | 0.383464789 | 1.185567285 | 231 |
| HP_ABNORMAL_SOCIAL_BEHAVIOR | 0.022980004 | 0.322177822 | 0.352487858 | 0.383916041 | 1.180532579 | 199 |
| LEIN_MIDBRAIN_MARKERS | 0.010835718 | 0.291209924 | 0.380730401 | 0.43222873 | 1.277970006 | 83 |
| KIM_ALL_DISORDERS_DURATION_CORR_DN | 0.000645701 | 0.149405595 | 0.477270815 | 0.44172103 | 1.335387542 | 139 |
| HP_INAPPROPRIATE_BEHAVIOR | 0.032967033 | 0.322177822 | 0.248911114 | 0.452530886 | 1.293058818 | 51 |
| GOBP_SENSORY_PERCEPTION_OF_TASTE | 0.027786217 | 0.322177822 | 0.352487858 | 0.457910717 | 1.310106644 | 52 |
| GOBP_MICROGLIAL_CELL_ACTIVATION | 0.031968032 | 0.322177822 | 0.252961123 | 0.458315847 | 1.298119984 | 43 |
| LEIN_PONS_MARKERS | 0.002016416 | 0.162573561 | 0.431707696 | 0.458467311 | 1.358726172 | 91 |
| WP_NEURODEGENERATION_WITH_BRAIN_IRON_ACCUMULATION_NBIA_SUBTYPES_PATHWAY | 0.031968032 | 0.322177822 | 0.252961123 | 0.45857068 | 1.299332761 | 44 |
| MCCLUNG_DELTA_FOSB_TARGETS_2WK | 0.013824839 | 0.307483498 | 0.380730401 | 0.472234124 | 1.348459241 | 50 |
| LEIN_OLIGODENDROCYTE_MARKERS | 0.00129778 | 0.149405595 | 0.455059867 | 0.477965072 | 1.405487697 | 77 |
| GOBP_LONG_TERM_MEMORY | 0.027972028 | 0.322177822 | 0.271288555 | 0.480561132 | 1.338666138 | 35 |
| REACTOME_DNA_DAMAGE_RECOGNITION_IN_GG_NER | 0.026284274 | 0.322177822 | 0.352487858 | 0.480715015 | 1.350414748 | 38 |
| GOBP_SENSORY_PERCEPTION_OF_BITTER_TASTE | 0.017272647 | 0.322177822 | 0.352487858 | 0.4979942 | 1.38439026 | 34 |
| HP_PROGRESSIVE_NEUROLOGIC_DETERIORATION | 0.002645375 | 0.18958518 | 0.431707696 | 0.504122292 | 1.439515547 | 50 |
| LEIN_LOCALIZED_TO_PROXIMAL_DENDRITES | 0.009640073 | 0.270341171 | 0.380730401 | 0.512685863 | 1.432533397 | 36 |
| JOHANSSON_GLIOMAGENESIS_BY_PDGFB_DN | 0.042296073 | 0.363746224 | 0.219250347 | 0.52270833 | 1.365622824 | 20 |
| HP_ADDICTIVE_BEHAVIOR | 0.029948776 | 0.322177822 | 0.352487858 | 0.539378548 | 1.435418228 | 23 |
| HP_ABNORMAL_METABOLIC_BRAIN_IMAGING_BY_MRS | 0.004796351 | 0.237972816 | 0.407017919 | 0.544114409 | 1.505110857 | 32 |
| GOBP_BEHAVIORAL_RESPONSE_TO_PAIN | 0.035425101 | 0.322177822 | 0.241339977 | 0.559469183 | 1.41982836 | 16 |

| pathway | pval | fdr | log2err | ES | NES | size |
|---|---|---|---|---|---|---|
| GOBP_AXON_ENSHEATHMENT_IN_CENTRAL_NERVOUS_SYSTEM | 0.007488302 | 0.270341171 | 0.407017919 | 0.58328874 | 1.552273987 | 23 |
| HP_ABNORMAL_BRAIN_LACTATE_LEVEL_BY_MRS | 0.015994415 | 0.322177822 | 0.352487858 | 0.601578599 | 1.51573933 | 15 |
| GOBP_EMBRYONIC_BRAIN_DEVELOPMENT | 0.007774827 | 0.270341171 | 0.407017919 | 0.607871761 | 1.567979871 | 18 |

Table S4 Selected radiation-specific significant gene sets after GSEA. for cerebellum after six months of radiation exposure.

| pathway | pval | fdr | log2err | ES | NES | size |
|---|---|---|---|---|---|---|
| BRUINS_UVC_RESPONSE_EARLY_LATE | 0.000271139 | 0.006319346 | 0.498493109 | 0.370990341 | 1.269016017 | 311 |
| BYSTRYKH_HEMATOPOIESIS_STEM_CELL_AND_BRAIN_QTL_TRANS | 0.010387351 | 0.056778316 | 0.380730401 | 0.373165752 | 1.237698937 | 159 |
| CHESLER_BRAIN_HIGHEST_EXPRESSION | 0.047952048 | 0.136854296 | 0.204294757 | 0.435301163 | 1.30342004 | 37 |
| DACOSTA_UV_RESPONSE_VIA_ERCC3_COMMON_DN | 1.41164E-13 | 9.10511E-11 | 0.943632225 | 0.430902957 | 1.491534499 | 461 |
| DACOSTA_UV_RESPONSE_VIA_ERCC3_TTD_DN | 0.004870712 | 0.03611045 | 0.407017919 | 0.428469493 | 1.369647111 | 84 |
| DACOSTA_UV_RESPONSE_VIA_ERCC3_UP | 0.01787342 | 0.07896134 | 0.352487858 | 0.341204711 | 1.166358442 | 306 |
| DACOSTA_UV_RESPONSE_VIA_ERCC3_XPCS_DN | 2.47137E-05 | 0.00122618 | 0.575610261 | 0.49286874 | 1.575505976 | 84 |
| DAZARD_RESPONSE_TO_UV_NHEK_DN | 5.88009E-09 | 1.26422E-06 | 0.761460801 | 0.427044817 | 1.456769478 | 296 |
| DAZARD_RESPONSE_TO_UV_SCC_DN | 0.006283566 | 0.042662105 | 0.407017919 | 0.399200891 | 1.302393483 | 115 |
| DAZARD_UV_RESPONSE_CLUSTER_G6 | 4.29365E-05 | 0.001730876 | 0.557332239 | 0.429934851 | 1.419757352 | 144 |
| ENK_UV_RESPONSE_EPIDERMIS_DN | 0.0001213 | 0.003556292 | 0.538434096 | 0.358704503 | 1.244715526 | 499 |
| ENK_UV_RESPONSE_EPIDERMIS_UP | 0.031968032 | 0.107392607 | 0.252961123 | 0.337161046 | 1.147919686 | 288 |
| ENK_UV_RESPONSE_KERATINOCYTE_DN | 5.91745E-07 | 9.29013E-05 | 0.659444398 | 0.381451985 | 1.321285678 | 468 |
| GENTILE_UV_HIGH_DOSE_DN | 0.000407845 | 0.007470507 | 0.498493109 | 0.371061094 | 1.262747279 | 284 |
| GENTILE_UV_RESPONSE_CLUSTER_D4 | 0.00464763 | 0.035687161 | 0.407017919 | 0.46423569 | 1.43638727 | 53 |
| GENTILE_UV_RESPONSE_CLUSTER_D6 | 0.045 | 0.135630841 | 0.211400189 | 0.447385237 | 1.324965816 | 34 |
| GENTILE_UV_RESPONSE_CLUSTER_D8 | 0.006209205 | 0.042662105 | 0.407017919 | 0.49594922 | 1.485018206 | 37 |
| GOBP_CELLULAR_RESPONSE_TO_GAMMA_RADIATION | 0.046 | 0.136728111 | 0.208955028 | 0.464445188 | 1.335493664 | 27 |
| GOBP_CELLULAR_RESPONSE_TO_IONIZING_RADIATION | 0.002127409 | 0.022494731 | 0.431707696 | 0.442534564 | 1.39834925 | 72 |
| GOBP_CEREBELLAR_CORTEX_FORMATION | 0.028028028 | 0.10099485 | 0.271288555 | 0.48839822 | 1.387460335 | 25 |
| GOBP_COGNITION | 0.002016416 | 0.022043873 | 0.431707696 | 0.359837824 | 1.229761333 | 301 |
| GOBP_CYTOPLASMIC_PATTERN_RECOGNITION_RECEPTOR_SIGNALING_PATHWAY | 0.001500267 | 0.018973967 | 0.455059867 | 0.467342864 | 1.462319447 | 61 |
| GOBP_CYTOPLASMIC_PATTERN_RECOGNITION_RECEPTOR_SIGNALING_PATHWAY_IN_RESPONSE_TO_VIRUS | 0.009350503 | 0.054172108 | 0.380730401 | 0.502796039 | 1.489069171 | 34 |
| GOBP_ERAD_PATHWAY | 0.017272647 | 0.076833498 | 0.352487858 | 0.384374022 | 1.247885044 | 107 |
| GOBP_HINDBRAIN_DEVELOPMENT | 0.037130842 | 0.116826308 | 0.321775918 | 0.362766785 | 1.198487424 | 147 |

| pathway | pval | fdr | log2err | ES | NES | size |
|---|---|---|---|---|---|---|
| GOBP_LEARNING | 0.047952048 | 0.136854296 | 0.204294757 | 0.355076007 | 1.172226993 | 143 |
| GOBP_MACROPHAGE_ACTIVATION_INVOLVED_IN_IMMUNE_RESPONSE | 0.038461538 | 0.120425691 | 0.23112671 | 0.5200985 | 1.399406851 | 17 |
| GOBP_MEMORY | 0.032300928 | 0.107392607 | 0.321775918 | 0.374483961 | 1.225646838 | 122 |
| GOBP_POSITIVE_REGULATION_OF_MACROAUTOPHAGY | 0.001702755 | 0.019892918 | 0.455059867 | 0.461026278 | 1.455724422 | 71 |
| GOBP_REGULATION_OF_DNA_TEMPLATED_TRANSCRIPTION_IN_RESPONSE_TO_STRESS | 0.013227015 | 0.064146048 | 0.380730401 | 0.442729132 | 1.369844034 | 53 |
| GOBP_REGULATION_OF_NEUROGENESIS | 0.005391236 | 0.039515312 | 0.407017919 | 0.34777062 | 1.194321317 | 350 |
| GOBP_REGULATION_OF_PATTERN_RECOGNITION_RECEPTOR_SIGNALING_PATHWAY | 0.040959041 | 0.125802769 | 0.222056046 | 0.376624256 | 1.218414989 | 101 |
| GOBP_REGULATION_OF_RIG_I_SIGNALING_PATHWAY | 0.01653669 | 0.075198315 | 0.352487858 | 0.53813601 | 1.483586683 | 20 |
| GOBP_REGULATION_OF_VIRAL_INDUCED_CYTOPLASMIC_PATTERN_RECOGNITION_RECEPTOR_SIGNALING_PATHWAY | 0.003176034 | 0.028062216 | 0.431707696 | 0.559999727 | 1.573069647 | 23 |
| GOBP_RESPONSE_TO_GAMMA_RADIATION | 0.010985174 | 0.059172644 | 0.380730401 | 0.446123647 | 1.38034697 | 53 |
| GOBP_RESPONSE_TO_IONIZING_RADIATION | 0.001334596 | 0.017216285 | 0.455059867 | 0.406627687 | 1.340760362 | 139 |
| GOBP_RESPONSE_TO_RADIATION | 0.007175897 | 0.044936441 | 0.407017919 | 0.339154782 | 1.171903241 | 424 |
| GOBP_RHYTHMIC_BEHAVIOR | 0.022980004 | 0.092638141 | 0.352487858 | 0.445302437 | 1.361399722 | 45 |
| GOBP_RIG_I_SIGNALING_PATHWAY | 0.01310374 | 0.064029637 | 0.380730401 | 0.499898858 | 1.444136067 | 28 |
| GOBP_SENSORY_PERCEPTION_OF_MECHANICAL_STIMULUS | 0.026973027 | 0.098291539 | 0.276500599 | 0.3582989 | 1.194262634 | 174 |
| GOBP_UBIQUITIN_DEPENDENT_ERAD_PATHWAY | 0.006581009 | 0.04376032 | 0.407017919 | 0.420699937 | 1.34481092 | 84 |
| GOCC_MYELIN_SHEATH | 0.040959041 | 0.125802769 | 0.222056046 | 0.421784901 | 1.284998867 | 44 |
| GOCC_RNA_POLYMERASE_II_HOLOENZYME | 0.028971029 | 0.102213115 | 0.266350657 | 0.40193399 | 1.266415431 | 69 |
| GSE29949_DC_BRAIN_VS_MONOCYTE_BONE_MARROW_UP | 0.000746448 | 0.011742894 | 0.477270815 | 0.385600842 | 1.290876419 | 196 |
| GSE29949_MICROGLIA_VS_DC_BRAIN_UP | 0.004796351 | 0.035972635 | 0.407017919 | 0.373244351 | 1.250112366 | 199 |
| HP_ABNORMALITY_OF_THE_BRAINSTEM_WHITE_MATTER | 0.025562372 | 0.097402597 | 0.287857117 | 0.549571795 | 1.446514732 | 15 |
| HP_ABNORMAL_AGGRESSIVE_IMPULSIVE_OR_VIOLENT_BEHAVIOR | 1.03532E-06 | 9.53975E-05 | 0.643551836 | 0.397432159 | 1.364385396 | 348 |
| HP_ABNORMAL_BRAINSTEM_MORPHOLOGY | 0.003089347 | 0.027675399 | 0.431707696 | 0.363459974 | 1.229072638 | 242 |
| HP_ABNORMAL_CONSUMPTION_BEHAVIOR | 0.001757979 | 0.019892918 | 0.455059867 | 0.40755845 | 1.341440846 | 134 |
| HP_ABNORMAL_EATING_BEHAVIOR | 0.000307594 | 0.006399932 | 0.498493109 | 0.442369267 | 1.429871128 | 100 |
| HP_ABNORMAL_FEAR_ANXIETY_RELATED_BEHAVIOR | 7.16323E-05 | 0.002310143 | 0.538434096 | 0.373713587 | 1.289569101 | 392 |
| HP_ABNORMAL_SOCIAL_BEHAVIOR | 0.016071103 | 0.074041866 | 0.352487858 | 0.360634326 | 1.207877436 | 199 |

| pathway | pval | fdr | log2err | ES | NES | size |
|---|---|---|---|---|---|---|
| HP_AGGRESSIVE_BEHAVIOR | 0.001113701 | 0.014965357 | 0.455059867 | 0.378347283 | 1.277176893 | 231 |
| HP_BENIGN_NEOPLASM_OF_THE_CENTRAL_NERVOUS_SYSTEM | 0.01262919 | 0.063639277 | 0.380730401 | 0.405498521 | 1.302079465 | 89 |
| HP_GLOBAL_BRAIN_ATROPHY | 0.046953047 | 0.136854296 | 0.206587923 | 0.406428934 | 1.262999011 | 55 |
| HP_INAPPROPRIATE_BEHAVIOR | 0.043956044 | 0.13373419 | 0.213927855 | 0.413096524 | 1.273331243 | 51 |
| HP_OBSESSIVE_COMPULSIVE_BEHAVIOR | 0.000284809 | 0.006319346 | 0.498493109 | 0.432321898 | 1.416857428 | 126 |
| HP_SELF_INJURIOUS_BEHAVIOR | 6.60175E-06 | 0.000473125 | 0.610526878 | 0.428069437 | 1.429807128 | 179 |
| HP_SPECIFIC_LEARNING_DISABILITY | 0.015169946 | 0.071945702 | 0.380730401 | 0.349567588 | 1.182496215 | 246 |
| JOHANSSON_BRAIN_CANCER_EARLY_VS_LATE_DN | 0.048951049 | 0.138479941 | 0.202071709 | 0.422060458 | 1.282799695 | 43 |
| KYNG_DNA_DAMAGE_BY_GAMMA_RADIATION | 0.04995005 | 0.139426901 | 0.199915231 | 0.387172494 | 1.226990279 | 75 |
| LEIN_CEREBELLUM_MARKERS | 0.006283566 | 0.042662105 | 0.407017919 | 0.414880753 | 1.327343477 | 85 |
| LEIN_LOCALIZED_TO_PROXIMAL_DENDRITES | 0.004721991 | 0.035831577 | 0.407017919 | 0.498999697 | 1.489541923 | 36 |
| LEIN_NEURON_MARKERS | 0.000412402 | 0.007470507 | 0.498493109 | 0.474313343 | 1.489483892 | 66 |
| LEIN_PONS_MARKERS | 0.024975025 | 0.096460426 | 0.287857117 | 0.389706086 | 1.252551103 | 91 |
| MCCLUNG_DELTA_FOSB_TARGETS_8WK | 0.019675741 | 0.08294675 | 0.352487858 | 0.436510452 | 1.341048426 | 49 |
| MEISSNER_BRAIN_ICP_WITH_H3K4ME3 | 0.002057525 | 0.02211839 | 0.431707696 | 0.539488366 | 1.564250279 | 29 |
| MONNIER_POSTRADIATION_TUMOR_ESCAPE_DN | 0.000100962 | 0.003100964 | 0.538434096 | 0.366469182 | 1.263092208 | 380 |
| MONNIER_POSTRADIATION_TUMOR_ESCAPE_UP | 1.82203E-11 | 5.87605E-09 | 0.863415392 | 0.428485608 | 1.479552326 | 397 |
| RASHI_RESPONSE_TO_IONIZING_RADIATION_4 | 0.006952814 | 0.04484565 | 0.407017919 | 0.457605198 | 1.405855755 | 49 |
| REACTOME_ACTIVATION_OF_ANTERIOR_HOX_GENES_IN_HINDBRAIN_DEVELOPMENT_DURING_EARLY_EMBRYOGENESIS | 0.01113463 | 0.059172644 | 0.380730401 | 0.401208544 | 1.295330572 | 97 |
| SESTO_RESPONSE_TO_UV_C2 | 0.008743341 | 0.051738117 | 0.380730401 | 0.445928717 | 1.385746637 | 55 |
| SESTO_RESPONSE_TO_UV_C3 | 0.045408678 | 0.136226034 | 0.211400189 | 0.493126595 | 1.359500266 | 20 |
| SESTO_RESPONSE_TO_UV_C5 | 0.031968032 | 0.107392607 | 0.252961123 | 0.423665274 | 1.297827473 | 46 |
| SESTO_RESPONSE_TO_UV_C8 | 0.009938984 | 0.054791834 | 0.380730401 | 0.433378855 | 1.368425649 | 71 |
| SIMBULAN_UV_RESPONSE_IMMORTALIZED_DN | 0.029 | 0.102213115 | 0.266350657 | 0.467432066 | 1.367999187 | 31 |
| WP_NEURODEGENERATION_WITH_BRAIN_IRON_ACCUMULATION_NBIA_SUBTYPES_PATHWAY | 0.019074967 | 0.081479162 | 0.352487858 | 0.447064966 | 1.362016456 | 44 |

*Table S5 Selected radiation-specific significant gene sets after GSEA. for cortex after one month of exposure to radiation.*

| pathway | pval | fdr | log2err | ES | NES | size |
|---|---|---|---|---|---|---|
| DACOSTA_UV_RESPONSE_VIA_ERCC3_COMMON_DN | 0.044955045 | 0.630347913 | 0.211400189 | 0.356827413 | 1.100482731 | 461 |
| DAZARD_RESPONSE_TO_UV_SCC_DN | 0.002978354 | 0.326548854 | 0.431707696 | 0.440936133 | 1.300033183 | 115 |
| DAZARD_UV_RESPONSE_CLUSTER_G28 | 0.010471035 | 0.557091029 | 0.380730401 | 0.619162354 | 1.533723059 | 16 |
| GENTILE_UV_RESPONSE_CLUSTER_D4 | 0.025974026 | 0.557091029 | 0.28201335 | 0.467146368 | 1.312828718 | 53 |
| GENTILE_UV_RESPONSE_CLUSTER_D9 | 0.026052104 | 0.557091029 | 0.28201335 | 0.524790232 | 1.363569623 | 25 |
| GHANDHI_BYSTANDER_IRRADIATION_UP | 0.000339492 | 0.093143577 | 0.498493109 | 0.500068527 | 1.436608011 | 76 |
| GHANDHI_DIRECT_IRRADIATION_UP | 0.017573033 | 0.557091029 | 0.352487858 | 0.435215811 | 1.266617492 | 93 |
| GOBP_FOREBRAIN_NEURON_DEVELOPMENT | 0.025100402 | 0.557091029 | 0.287857117 | 0.54081297 | 1.388069684 | 22 |
| GOBP_GAMMA_DELTA_T_CELL_ACTIVATION | 0.013019145 | 0.557091029 | 0.380730401 | 0.579079088 | 1.459811167 | 19 |
| GOBP_LEARNING | 0.037962038 | 0.569430569 | 0.23112671 | 0.396338914 | 1.18115404 | 143 |
| GOBP_LONG_TERM_MEMORY | 0.02997003 | 0.557091029 | 0.261663522 | 0.491784637 | 1.331184616 | 35 |
| GOBP_RESPONSE_TO_GAMMA_RADIATION | 0.030969031 | 0.557091029 | 0.257206466 | 0.457779467 | 1.286504772 | 53 |
| GOBP_VOCALIZATION_BEHAVIOR | 0.000231981 | 0.093143577 | 0.518848078 | 0.71582526 | 1.773166118 | 16 |
| GOCC_AMPA_GLUTAMATE_RECEPTOR_COMPLEX | 0.003543941 | 0.326548854 | 0.431707696 | 0.584896393 | 1.524646224 | 26 |
| GOCC_TRANSCRIPTION_FACTOR_TFIID_COMPLEX | 0.040959041 | 0.600422305 | 0.222056046 | 0.495278316 | 1.319607487 | 31 |
| KYNG_ENVIRONMENTAL_STRESS_RESPONSE_NOT_BY_GAMMA_IN_OLD | 0.011606187 | 0.557091029 | 0.380730401 | 0.542103577 | 1.425744869 | 28 |
| MCCLUNG_COCAIN_REWARD_4WK | 0.012330278 | 0.557091029 | 0.380730401 | 0.447288542 | 1.284980494 | 76 |
| MONNIER_POSTRADIATION_TUMOR_ESCAPE_DN | 0.006283566 | 0.506612492 | 0.407017919 | 0.380380459 | 1.170276835 | 380 |
| MURAKAMI_UV_RESPONSE_24HR | 0.043259557 | 0.620053655 | 0.216542837 | 0.537918987 | 1.356049906 | 19 |
| RASHI_RESPONSE_TO_IONIZING_RADIATION_1 | 0.026973027 | 0.557091029 | 0.276500599 | 0.484179334 | 1.321040358 | 39 |

*Table S6 Selected radiation-specific significant gene sets after GSEA. for cortex after six months of exposure to radiation.*

| pathway | pval | fdr | log2err | ES | NES | size |
|---|---|---|---|---|---|---|
| BRUINS_UVC_RESPONSE_EARLY_LATE | 0.006804092 | 0.091288235 | 0.407017919 | 0.353000782 | 1.193173679 | 311 |
| BRUINS_UVC_RESPONSE_VIA_TP53_GROUP_C | 0.026284274 | 0.180339771 | 0.352487858 | 0.395983818 | 1.256561353 | 89 |
| DACOSTA_UV_RESPONSE_VIA_ERCC3_COMMON_DN | 3.53725E-07 | 0.000227799 | 0.67496286 | 0.390540954 | 1.33827574 | 461 |
| DAZARD_RESPONSE_TO_UV_NHEK_DN | 9.15535E-07 | 0.000294802 | 0.659444398 | 0.410227435 | 1.385143513 | 296 |
| DAZARD_RESPONSE_TO_UV_SCC_DN | 0.000966438 | 0.033447929 | 0.477270815 | 0.42819065 | 1.38518469 | 115 |
| DAZARD_UV_RESPONSE_CLUSTER_G6 | 0.008145521 | 0.097142875 | 0.380730401 | 0.384522296 | 1.259441504 | 144 |
| ENK_UV_RESPONSE_EPIDERMIS_DN | 0.002386391 | 0.054887001 | 0.431707696 | 0.342905205 | 1.17811725 | 499 |
| ENK_UV_RESPONSE_KERATINOCYTE_DN | 0.005391236 | 0.081938706 | 0.407017919 | 0.340444482 | 1.167186654 | 468 |
| GENTILE_UV_HIGH_DOSE_DN | 0.007547702 | 0.097142875 | 0.407017919 | 0.356347149 | 1.201601776 | 284 |
| GENTILE_UV_RESPONSE_CLUSTER_D8 | 0.004503408 | 0.074363972 | 0.407017919 | 0.506500349 | 1.486070834 | 37 |
| GOBP_CELL_PROLIFERATION_IN_FOREBRAIN | 0.004890253 | 0.078733079 | 0.407017919 | 0.567621825 | 1.570352712 | 23 |
| GOBP_CENTRAL_NERVOUS_SYSTEM_NEURON_AXONOGENESIS | 0.017891293 | 0.15059975 | 0.352487858 | 0.480064072 | 1.400488647 | 35 |
| GOBP_CENTRAL_NERVOUS_SYSTEM_NEURON_DIFFERENTIATION | 0.014721577 | 0.142939488 | 0.380730401 | 0.371196329 | 1.225705731 | 169 |
| GOBP_CENTRAL_NERVOUS_SYSTEM_PROJECTION_NEURON_AXONOGENESIS | 0.03109328 | 0.189476482 | 0.257206466 | 0.501224378 | 1.410295159 | 26 |
| GOBP_IMMUNOLOGICAL_MEMORY_PROCESS | 0.038500507 | 0.206619385 | 0.23112671 | 0.535797672 | 1.41355741 | 17 |
| GOBP_MIDBRAIN_DOPAMINERGIC_NEURON_DIFFERENTIATION | 0.035460993 | 0.200323504 | 0.241339977 | 0.542928398 | 1.432369904 | 17 |
| GOBP_NATURAL_KILLER_CELL_MEDIATED_IMMUNITY | 0.032967033 | 0.196777778 | 0.248911114 | 0.407597607 | 1.270516791 | 69 |
| GOBP_NEURAL_TUBE_FORMATION | 0.035964036 | 0.201398601 | 0.237793834 | 0.383588433 | 1.231150133 | 103 |
| GOBP_OLFACTORY_LOBE_DEVELOPMENT | 0.04 | 0.209430894 | 0.224966094 | 0.462988992 | 1.350675597 | 35 |
| GOBP_POSITIVE_REGULATION_OF_MACROAUTOPHAGY | 0.014422665 | 0.142895323 | 0.380730401 | 0.42612352 | 1.331785902 | 71 |
| GOBP_RESPONSE_TO_LIGHT_STIMULUS | 0.007697156 | 0.097142875 | 0.407017919 | 0.350962919 | 1.185303232 | 298 |
| HP_ABNORMAL_AGGRESSIVE_IMPULSIVE_OR_VIOLENT_BEHAVIOR | 0.001095293 | 0.035268438 | 0.455059867 | 0.362416739 | 1.230726326 | 348 |
| HP_ABNORMAL_BRAINSTEM_MORPHOLOGY | 0.016371489 | 0.148496319 | 0.352487858 | 0.355349124 | 1.190302725 | 242 |
| HP_ABNORMAL_BRAINSTEM_MRI_SIGNAL_INTENSITY | 0.008004062 | 0.097142875 | 0.380730401 | 0.500338031 | 1.460516032 | 36 |
| HP_ABNORMAL_BRAIN_POSITRON_EMISSION_TOMOGRAPHY | 0.030090271 | 0.188137227 | 0.261663522 | 0.507406577 | 1.40376437 | 23 |

| pathway | pval | fdr | log2err | ES | NES | size |
|---|---|---|---|---|---|---|
| HP_ABNORMAL_CNS_MYELINATION | 0.00029848 | 0.014786237 | 0.498493109 | 0.378836852 | 1.279058939 | 293 |
| HP_ABNORMAL_FEAR_ANXIETY_RELATED_BEHAVIOR | 0.010237895 | 0.109886741 | 0.380730401 | 0.342120254 | 1.165860549 | 392 |
| HP_ABNORMAL_SOCIAL_BEHAVIOR | 3.90077E-05 | 0.00448803 | 0.557332239 | 0.414136515 | 1.375934785 | 199 |
| HP_DELAYED_CNS_MYELINATION | 0.030969031 | 0.189476482 | 0.257206466 | 0.396260282 | 1.251795291 | 80 |
| HP_FOCAL_T2_HYPERINTENSE_BRAINSTEM_LESION | 0.002502384 | 0.05557018 | 0.431707696 | 0.544420023 | 1.548526841 | 29 |
| HP_SELF_INJURIOUS_BEHAVIOR | 0.001353004 | 0.041492111 | 0.455059867 | 0.393480647 | 1.303086053 | 179 |
| JOHANSSON_GLIOMAGENESIS_BY_PDGFB_DN | 0.031187123 | 0.189476482 | 0.257206466 | 0.51727529 | 1.400449757 | 20 |
| LEIN_LOCALIZED_TO_DISTAL_AND_PROXIMAL_DENDRITES | 0.004261063 | 0.072464035 | 0.407017919 | 0.603600011 | 1.592435564 | 17 |
| LEIN_NEURON_MARKERS | 0.000139438 | 0.009977542 | 0.518848078 | 0.49652346 | 1.54316729 | 66 |
| LIU_IL13_MEMORY_MODEL_UP | 0.026342452 | 0.180339771 | 0.28201335 | 0.554069655 | 1.461763101 | 17 |
| MONNIER_POSTRADIATION_TUMOR_ESCAPE_DN | 0.000293923 | 0.014786237 | 0.498493109 | 0.368489433 | 1.254467253 | 380 |
| MONNIER_POSTRADIATION_TUMOR_ESCAPE_UP | 5.3039E-05 | 0.00487959 | 0.557332239 | 0.373821079 | 1.274385617 | 397 |
| MURAKAMI_UV_RESPONSE_24HR | 0.005429203 | 0.081938706 | 0.407017919 | 0.575433974 | 1.546504657 | 19 |
| MURAKAMI_UV_RESPONSE_6HR_DN | 0.027162978 | 0.180339771 | 0.276500599 | 0.529295849 | 1.432993722 | 20 |
| RASHI_RESPONSE_TO_IONIZING_RADIATION_4 | 0.03996004 | 0.209430894 | 0.224966094 | 0.429306729 | 1.295927909 | 49 |
| REACTOME_CARGO_RECOGNITION_FOR_CLATHRIN_MEDIATED_ENDOCYTOSIS | 0.008743341 | 0.098740699 | 0.380730401 | 0.405484948 | 1.301428312 | 103 |
| REACTOME_DNA_DAMAGE_RECOGNITION_IN_GG_NER | 0.033 | 0.196777778 | 0.248911114 | 0.464556569 | 1.366295575 | 38 |
| SESTO_RESPONSE_TO_UV_C1 | 0.003570318 | 0.067626016 | 0.431707696 | 0.449078398 | 1.401601395 | 70 |
| SESTO_RESPONSE_TO_UV_C2 | 0.006432288 | 0.088136025 | 0.407017919 | 0.461092224 | 1.406398961 | 55 |

*Table S7 Selected radiation-specific significant gene sets after GSEA. for hippocampus after one month of exposure to radiation.*

| pathway | pval | fdr | log2err | ES | NES | size |
|---|---|---|---|---|---|---|
| DAZARD_UV_RESPONSE_CLUSTER_G4 | 0.039314516 | 1 | 0.227987203 | 0.54322928 | 1.369222262 | 19 |
| GOBP_APOPTOTIC_CELL_CLEARANCE | 0.007697156 | 0.883816287 | 0.407017919 | 0.494275389 | 1.369401643 | 48 |
| GOBP_GROOMING_BEHAVIOR | 0.009454503 | 0.883816287 | 0.380730401 | 0.593890009 | 1.484238062 | 18 |
| GOBP_MACROPHAGE_ACTIVATION_INVOLVED_IN_IMMUNE_RESPONSE | 0.031376518 | 1 | 0.257206466 | 0.564387605 | 1.396133714 | 17 |
| GOBP_NATURAL_KILLER_CELL_ACTIVATION | 0.004275828 | 0.687339373 | 0.407017919 | 0.459674377 | 1.326670042 | 84 |
| GOBP_POSITIVE_REGULATION_OF_CIRCADIAN_RHYTHM | 0.041497976 | 1 | 0.222056046 | 0.555885413 | 1.375101718 | 17 |
| GOBP_RESPONSE_TO_DIETARY_EXCESS | 0.010846554 | 0.883816287 | 0.380730401 | 0.528741455 | 1.400012916 | 29 |
| GOBP_RHYTHMIC_BEHAVIOR | 0.016371489 | 1 | 0.352487858 | 0.482279255 | 1.329943237 | 45 |
| HP_ABNORMALITY_OF_THE_BRAINSTEM_WHITE_MATTER | 0.024219908 | 1 | 0.352487858 | 0.591988633 | 1.435800129 | 15 |
| MCCLUNG_COCAIN_REWARD_4WK | 0.026973027 | 1 | 0.276500599 | 0.437576454 | 1.253250481 | 76 |
| SIMBULAN_UV_RESPONSE_IMMORTALIZED_DN | 0.010996159 | 0.883816287 | 0.380730401 | 0.530155847 | 1.414825892 | 31 |
| TSAI_RESPONSE_TO_RADIATION_THERAPY | 0.002536916 | 0.687339373 | 0.431707696 | 0.56437291 | 1.506140902 | 31 |

Table S8 Selected radiation-specific significant gene sets after GSEA. for hippocampus after six months of exposure to radiation.

| pathway | pval | fdr | log2err | ES | NES | size |
|---|---|---|---|---|---|---|
| DAZARD_UV_RESPONSE_CLUSTER_G6 | 0.007846611 | 0.514946629 | 0.380730401 | 0.423817591 | 1.241462689 | 144 |
| ENK_UV_RESPONSE_KERATINOCYTE_DN | 0.000221058 | 0.142361542 | 0.518848078 | 0.399948108 | 1.210695732 | 468 |
| GENTILE_RESPONSE_CLUSTER_D3 | 0.02718544 | 1 | 0.352487858 | 0.465746079 | 1.288429696 | 53 |
| GENTILE_UV_HIGH_DOSE_DN | 0.040959041 | 1 | 0.222056046 | 0.380334809 | 1.136862596 | 284 |
| GENTILE_UV_HIGH_DOSE_UP | 0.040160643 | 1 | 0.224966094 | 0.53220722 | 1.35525562 | 22 |
| GENTILE_UV_RESPONSE_CLUSTER_D4 | 0.026885051 | 1 | 0.352487858 | 0.467645014 | 1.293682869 | 53 |
| GOBP_CELLULAR_RESPONSE_TO_IONIZING_RADIATION | 0.007996066 | 0.514946629 | 0.380730401 | 0.471657789 | 1.335865608 | 72 |
| GOBP_CELLULAR_RESPONSE_TO_RADIATION | 0.01787342 | 0.959206852 | 0.352487858 | 0.405047888 | 1.196492849 | 178 |
| GOBP_ERAD_PATHWAY | 0.002534382 | 0.441264322 | 0.431707696 | 0.460072383 | 1.330703841 | 107 |
| GOBP_MIDBRAIN_DEVELOPMENT | 0.037130842 | 1 | 0.321775918 | 0.434458291 | 1.241346715 | 87 |
| GOBP_MYELOID_DENDRITIC_CELL_ACTIVATION | 0.037 | 1 | 0.234392647 | 0.496545105 | 1.306505521 | 29 |
| GOBP_UBIQUITIN_DEPENDENT_ERAD_PATHWAY | 0.002127409 | 0.441264322 | 0.431707696 | 0.476432458 | 1.358182644 | 84 |
| GOCC_RNA_POLYMERASE_II_HOLOENZYME | 0.024975025 | 1 | 0.287857117 | 0.445101148 | 1.257470378 | 69 |
| HP_ABNORMAL_BRAINSTEM_MORPHOLOGY | 0.047952048 | 1 | 0.204294757 | 0.383341358 | 1.140753604 | 242 |
| HP_ATROPHY_DEGENERATION_AFFECTING_THE_BRAINSTEM | 0.022424032 | 1 | 0.352487858 | 0.537227255 | 1.391766378 | 25 |
| HP_HYPOPLASIA_OF_THE_BRAINSTEM | 0.004127107 | 0.441264322 | 0.407017919 | 0.47206717 | 1.341975646 | 77 |
| SESTO_RESPONSE_TO_UV_C0 | 0.025683497 | 1 | 0.352487858 | 0.418812427 | 1.215027477 | 117 |
| SMIRNOV_RESPONSE_TO_IR_2HR_UP | 0.004498909 | 0.441264322 | 0.407017919 | 0.504621653 | 1.393231427 | 52 |